\documentclass[11pt,a4paper]{article}

\usepackage[utf8]{inputenc}
\usepackage[T1]{fontenc}
\usepackage[margin=1in]{geometry}
\usepackage{amsmath}
\usepackage{amssymb}
\usepackage{stmaryrd}
\usepackage{booktabs}
\usepackage{array}
\usepackage{tabularx}
\usepackage{authblk}
\usepackage{xcolor}
\usepackage{graphicx}
\usepackage[export]{adjustbox}
\usepackage{longtable}
\usepackage{calc}
\usepackage{newunicodechar}
\usepackage{listings}
\usepackage[colorlinks=true,linkcolor=blue!40!black,citecolor=blue!40!black,urlcolor=blue!40!black]{hyperref}

\let\origtextunderscore\_
\renewcommand{\_}{\origtextunderscore\allowbreak}

\lstdefinestyle{leanbox}{
  basicstyle=\footnotesize\ttfamily,
  columns=fullflexible,
  keepspaces=true,
  breaklines=true,
  breakatwhitespace=false,
  showstringspaces=false,
  upquote=true,
  extendedchars=true,
  inputencoding=utf8,
  literate=
    {§}{{\S}}1
    {¬}{{\ensuremath{\neg}}}1
    {·}{{\ensuremath{\cdot}}}1
    {×}{{\ensuremath{\times}}}1
    {Γ}{{\ensuremath{\Gamma}}}1
    {Δ}{{\ensuremath{\Delta}}}1
    {Η}{{\ensuremath{H}}}1
    {Θ}{{\ensuremath{\Theta}}}1
    {Κ}{{\ensuremath{K}}}1
    {Σ}{{\ensuremath{\Sigma}}}1
    {α}{{\ensuremath{\alpha}}}1
    {β}{{\ensuremath{\beta}}}1
    {λ}{{\ensuremath{\lambda}}}1
    {σ}{{\ensuremath{\sigma}}}1
    {τ}{{\ensuremath{\tau}}}1
    {φ}{{\ensuremath{\varphi}}}1
    {χ}{{\ensuremath{\chi}}}1
    {ψ}{{\ensuremath{\psi}}}1
    {ᵢ}{{\textsubscript{i}}}1
    {—}{{\textemdash}}1
    {⁺}{{\textsuperscript{+}}}1
    {⁻}{{\textsuperscript{\ensuremath{-}}}}1
    {₀}{{\textsubscript{0}}}1
    {₁}{{\textsubscript{1}}}1
    {₂}{{\textsubscript{2}}}1
    {₃}{{\textsubscript{3}}}1
    {₊}{{\textsubscript{+}}}1
    {ₙ}{{\textsubscript{n}}}1
    {ₚ}{{\textsubscript{p}}}1
    {ℕ}{{\ensuremath{\mathbb{N}}}}1
    {←}{{\ensuremath{\leftarrow}}}1
    {↑}{{\ensuremath{\uparrow}}}1
    {→}{{\ensuremath{\rightarrow}}}1
    {↓}{{\ensuremath{\downarrow}}}1
    {↔}{{\ensuremath{\leftrightarrow}}}1
    {↥}{{\ensuremath{\Uparrow}}}1
    {↦}{{\ensuremath{\mapsto}}}1
    {⇒}{{\ensuremath{\Rightarrow}}}1
    {∀}{{\ensuremath{\forall}}}1
    {∃}{{\ensuremath{\exists}}}1
    {∅}{{\ensuremath{\emptyset}}}1
    {∈}{{\ensuremath{\in}}}1
    {∉}{{\ensuremath{\notin}}}1
    {∘}{{\ensuremath{\circ}}}1
    {∧}{{\ensuremath{\wedge}}}1
    {∨}{{\ensuremath{\vee}}}1
    {∪}{{\ensuremath{\cup}}}1
    {∼}{{\ensuremath{\sim}}}1
    {≠}{{\ensuremath{\neq}}}1
    {≤}{{\ensuremath{\leq}}}1
    {≥}{{\ensuremath{\geq}}}1
    {⊆}{{\ensuremath{\subseteq}}}1
    {⊢}{{\ensuremath{\vdash}}}1
    {⊥}{{\ensuremath{\bot}}}1
    {⊨}{{\ensuremath{\vDash}}}1
    {⊬}{{\ensuremath{\nvdash}}}1
    {⊭}{{\ensuremath{\nvDash}}}1
    {⋀}{{\ensuremath{\bigwedge}}}1
    {⋁}{{\ensuremath{\bigvee}}}1
    {─}{{\ensuremath{-}}}1
    {│}{{\ensuremath{|}}}1
    {└}{{+}}1
    {├}{{+}}1
    {□}{{\ensuremath{\square}}}1
    {▸}{{\ensuremath{\blacktriangleright}}}1
    {◇}{{\ensuremath{\lozenge}}}1
    {⟨}{{\ensuremath{\langle}}}1
    {⟩}{{\ensuremath{\rangle}}}1
    {⟶}{{\ensuremath{\longrightarrow}}}2
    {⟷}{{\ensuremath{\longleftrightarrow}}}2
    {⟹}{{\ensuremath{\Longrightarrow}}}2
}
\newunicodechar{α}{\ensuremath{\alpha}}
\newunicodechar{β}{\ensuremath{\beta}}
\newunicodechar{σ}{\ensuremath{\sigma}}
\newunicodechar{φ}{\ensuremath{\varphi}}
\newunicodechar{χ}{\ensuremath{\chi}}
\newunicodechar{ψ}{\ensuremath{\psi}}
\newunicodechar{Γ}{\ensuremath{\Gamma}}
\newunicodechar{Δ}{\ensuremath{\Delta}}
\newunicodechar{ℕ}{\ensuremath{\mathbb{N}}}
\newunicodechar{∀}{\ensuremath{\forall}}
\newunicodechar{∃}{\ensuremath{\exists}}
\newunicodechar{¬}{\ensuremath{\neg}}
\newunicodechar{∧}{\ensuremath{\wedge}}
\newunicodechar{□}{\ensuremath{\square}}
\newunicodechar{◇}{\ensuremath{\lozenge}}
\newunicodechar{∼}{\ensuremath{\sim}}
\newunicodechar{→}{\ensuremath{\rightarrow}}
\newunicodechar{←}{\ensuremath{\leftarrow}}
\newunicodechar{↔}{\ensuremath{\leftrightarrow}}
\newunicodechar{↦}{\ensuremath{\mapsto}}
\newunicodechar{⇒}{\ensuremath{\Rightarrow}}
\newunicodechar{⟶}{\ensuremath{\longrightarrow}}
\newunicodechar{⟷}{\ensuremath{\longleftrightarrow}}
\newunicodechar{⟹}{\ensuremath{\Longrightarrow}}
\newunicodechar{⟺}{\ensuremath{\Longleftrightarrow}}
\newunicodechar{∈}{\ensuremath{\in}}
\newunicodechar{⊢}{\ensuremath{\vdash}}
\newunicodechar{⊨}{\ensuremath{\vDash}}
\newunicodechar{≠}{\ensuremath{\neq}}
\newunicodechar{≈}{\ensuremath{\approx}}
\newunicodechar{≅}{\ensuremath{\cong}}
\newunicodechar{⊆}{\ensuremath{\subseteq}}
\newunicodechar{⊇}{\ensuremath{\supseteq}}
\newunicodechar{⊋}{\ensuremath{\supsetneq}}
\newunicodechar{⊥}{\ensuremath{\bot}}
\newunicodechar{∪}{\ensuremath{\cup}}
\newunicodechar{⊕}{\ensuremath{\oplus}}
\newunicodechar{×}{\ensuremath{\times}}
\newunicodechar{⟨}{\ensuremath{\langle}}
\newunicodechar{⟩}{\ensuremath{\rangle}}
\newunicodechar{₀}{\textsubscript{0}}
\newunicodechar{₁}{\textsubscript{1}}
\newunicodechar{ₙ}{\textsubscript{n}}
\newunicodechar{⁺}{\textsuperscript{+}}
\newunicodechar{·}{\textperiodcentered{}}
\newunicodechar{—}{\textemdash{}}
\newunicodechar{–}{\textendash{}}
\newunicodechar{…}{\ldots}
\newunicodechar{§}{\S}
\newunicodechar{│}{\ensuremath{\mid}}
\newunicodechar{✓}{\ensuremath{\checkmark}}
\newunicodechar{æ}{\ae}
\newunicodechar{ö}{\"o}

\title{\textbf{Finishing Oltean's Completeness Proof in Lean 4 for Hybrid Logic $L(\forall)$}}

        \author[1]{\textbf{Lars Warren Ericson}}
        \affil[1]{Catskills Research Company}
        \affil[1]{\texttt{lars.ericson@catskillsresearch.com}}

        \date{\today}

\begin{document}

        \maketitle

        \begin{center}
          \small
          \textbf{ORCID:} 0000-0001-8299-9361 \\
          \textbf{Primary Category:} cs.LO (Logic in Computer Science) \\
          \textbf{Secondary Category:} math.LO (Logic) \\[0.5em]
          \textbf{Lean~4 formalization:} \url{https://github.com/catskillsresearch/hybrid\_logic\_lean\_revisited}
        \end{center}

        \begin{abstract}
        We present a machine-checked completeness theorem, in \textbf{Lean 4}, for the hybrid logic \emph{L(∀)} --- propositional modal logic enriched with nominals, the satisfaction-style universal binder ∀, and the box modality. (Machine-checked completeness for \emph{basic} hybrid logic --- nominals and satisfaction operators, without binders --- was pioneered by Asta Halkjær From in Isabelle/HOL {[}Fro20{]}.) We build directly on Alex Oltean's 2023 Lean 4 formalization {[}Olt23{]}. Oltean mechanized the syntax, semantics, Hilbert-style proof system, and \textbf{soundness} following Blackburn's \emph{Hybrid Completeness} (1998) {[}Bla98{]} and laid out a clear route to completeness, but left the theorem itself unfinished. Finishing it requires manufacturing fresh names at two structurally different points of the proof, and our central finding is that \textbf{the two points call for two different tools}. \emph{(1) The root witnessed maximal consistent set} --- the extended Lindenbaum construction --- needs, at each step, a nominal fresh for the whole set being built; the right tool is \emph{structural freshness}, extending the language so that an infinite supply of nominals is reserved \emph{by construction} and is automatically disjoint from anything in play. We survey the design space for this --- Oltean's odd/even encoding inside ℕ, the disjoint-sum (\texttt{N\ ⊕\ ℕ}) parameterization suggested by Bud Mishra, and Asta Halkjær From's abstract synthetic-completeness frameworks --- and explain the encoding we adopt. \emph{(2) The witnessed ◇-successor} of a maximal consistent set, by contrast, \textbf{cannot} be obtained this way: its canonical box-reduct provably mentions every nominal, so no reserved name is ever fresh for it. Here the correct tool is the one Oltean had already chosen but left incomplete --- an \emph{existence-lemma} Henkin construction that draws each witness from the predecessor's \emph{own} witnessedness through a fresh \emph{state variable} rather than a fresh nominal; we complete it with a data-carrying witness accumulator and a compactness argument. With both constructions in place the completeness theorem \texttt{Γ\ ⊨\ φ\ →\ Γ\ ⊢\ φ} is fully formalized: the development is \texttt{sorry}-free, and \texttt{\#print\ axioms\ Completeness} reports only \texttt{propext}, \texttt{Classical.choice}, and \texttt{Quot.sound}. We also port the development from Oltean's original June-2023 Lean nightly to Lean v4.30.0 / mathlib v4.30.0.

\begin{center}\rule{0.5\linewidth}{0.5pt}\end{center}
        \end{abstract}

\hypertarget{introduction}{%
\section{Introduction}\label{introduction}}

\hypertarget{hybrid-logic}{%
\subsection{Hybrid logic}\label{hybrid-logic}}

Modal logic extends propositional logic with operators □ (``necessarily'') and ◇ (``possibly'') interpreted over Kripke frames --- directed graphs of ``states'' or ``worlds''. \emph{Hybrid} logic, originating in Arthur Prior's work on the logic of time, augments modal logic with \textbf{nominals}: atomic symbols \texttt{i,\ j,\ k,\ …} each true at \emph{exactly one} state, so that a nominal acts as a \emph{name} for that state. This modest addition dramatically increases expressive power while retaining good logical behavior, and it makes hybrid languages natural for talking about relational structures --- a perspective that has made them attractive for, e.g., XML constraints, description logics, and the relationship to Matching Logic and the K framework.

The system formalized here is \emph{L(∀)} (equivalently written \emph{H(∀)}): propositional hybrid logic with nominals, the box □, and the binder ∀x, where state variables \texttt{x} are simultaneously bindable variables and well-formed formulas. \texttt{∀x\ φ} quantifies over states; \texttt{∃x\ φ} abbreviates \texttt{¬∀x¬φ}. (This is the ``strong'' hybrid language with binding, as opposed to the weaker language whose only hybrid primitive is the satisfaction operator \texttt{@\_i}.)

\hypertarget{soundness-completeness-and-what-was-left-open}{%
\subsection{Soundness, completeness, and what was left open}\label{soundness-completeness-and-what-was-left-open}}

Oltean's formalization (\texttt{oltean\_thesis.pdf}; repository archived at \texttt{github.com/alexoltean61/hybrid\_logic\_lean}) defines:

\begin{itemize}
\item
  the syntax of \emph{L(∀)} and substitution machinery (\texttt{Form.lean}, \texttt{Substitutions.lean});
\item
  a Kripke semantics (\texttt{Truth.lean});
\item
  a Hilbert-style proof system (\texttt{Proof.lean});
\item
  and a proof of \textbf{soundness}, \texttt{Γ\ ⊢\ φ\ ⟹\ Γ\ ⊨\ φ} (\texttt{Soundness.lean}).
\end{itemize}

The converse, \textbf{completeness} (\texttt{Γ\ ⊨\ φ\ ⟹\ Γ\ ⊢\ φ}), was left as an open formalization problem. Oltean had already written much of the scaffolding --- the Lindenbaum construction, a notion of \emph{witnessed} set, the canonical/completed model, and the statements of the extended Lindenbaum lemma, the existence lemma, and the truth lemma --- but a number of key lemmas remained as \texttt{sorry}/\texttt{admit} placeholders.

\hypertarget{an-anecdote-henkin-mishra-and-the-shape-of-the-difficulty}{%
\subsection{An anecdote: Henkin, Mishra, and the shape of the difficulty}\label{an-anecdote-henkin-mishra-and-the-shape-of-the-difficulty}}

Why was completeness left open at all, when the textbook proof is routine? The answer is a small but instructive collision between classical mathematics and type theory, and it is worth telling as motivation.

The completeness proof is a \emph{Henkin construction}: one extends a consistent set to a maximal consistent set that is moreover \emph{witnessed} --- every existential \texttt{∃x\ φ} comes with a nominal \texttt{i} certifying it, \texttt{(∃x\ φ\ →\ φ{[}i/x{]})}. Each existential needs its \emph{own} witness, and to saturate an infinite set one needs an infinite reserve of nominals that do not already occur anywhere in play. In ordinary set-theoretic practice this is a non-issue: one simply says ``let \texttt{i₀,\ i₁,\ …} enumerate fresh nominals,'' because there is never a shortage of names. In Lean's dependent type theory the same sentence has no referent: a type \texttt{N} already contains \emph{all} of its inhabitants, and there is in general no \texttt{N\textquotesingle{}\ ⊋\ N} to draw new names from. Oltean could search a single formula for an unused nominal --- finitely many occur --- but witnessing the \emph{infinite} Lindenbaum union by repeated dynamic search turned out to be, in his words, prohibitively difficult. That is precisely where the formalization stalled.

When we set out to revive the (by then archived) development, the natural first question was whether this obstacle was fundamental --- whether the ``easy'' textbook proof was simply not available in type theory, and whether one ought instead to adopt a heavier, more abstract machine, such as the transfinite synthetic-completeness frameworks that Asta Halkjær From has developed in Isabelle/HOL. The decisive nudge came anecdotally. In discussions around the problem, \textbf{Bud Mishra} suggested the remedy that, in hindsight, is the canonical one: do not \emph{search} for fresh names --- \emph{reserve} them structurally. Parameterize formulas by their nominal type \texttt{N} and, when it is time to run Lindenbaum, pass to the disjoint sum \texttt{N\ ⊕\ ℕ}, drawing every Henkin witness from the right summand \texttt{Sum.inr\ n}. Freshness then ceases to be a computation and becomes a fact of the sum type: a witness is distinct from every base nominal because it lives in a different injection. This is exactly Henkin's old idea {[}Hen49{]} --- expand the language with new constants --- rendered in a form that type theory accepts without complaint.

Two realizations followed. First, \textbf{Oltean had already built this idea into his development}, in disguise: his \texttt{Form.odd\_noms} remaps every nominal \texttt{i\ ↦\ 2·i+1}, so that the image uses only odd nominals and \emph{all even nominals} are reserved as a fresh supply. The odd/even split inside ℕ is precisely \texttt{N\ ⊕\ ℕ} internalized (odds ≅ \texttt{Sum.inl}, evens ≅ \texttt{Sum.inr}). Indeed, From's Isabelle approach --- a fixed name type with a \texttt{fresh} operator returning an unused name --- is the same principle a third time over. So the structural-freshness idea is not a clever trick belonging to any one of these treatments; it is the shared, and essentially unavoidable, foundation of all of them. In that sense \textbf{Mishra's suggestion was not a ``bust,'' and there is no need to pivot wholesale to a Halkjær-style framework} to finish this particular proof: the right idea was already on the table, twice.

Second, and more usefully, the realization reframes \emph{where the real difficulty lies}. It is not in the freshness principle but in its \textbf{encoding}. Oltean implements the odd/even remapping as \texttt{bulk\_subst} --- an iterated single-nominal substitution walked in lockstep over the formula's list of nominals --- and that list, for a compound formula, is a \emph{merged, deduplicated, sorted} list rather than the concatenation of its parts' lists. Consequently the apparently trivial homomorphism lemma \texttt{(φ\ →\ ψ).odd\_noms\ =\ φ.odd\_noms\ →\ ψ.odd\_noms} becomes a genuine fight with ordering and deduplication, and every later step (theorem-preservation under expansion, ``enough nominals,'' the witnessed Lindenbaum lemma) waits on it. The lesson --- which we develop in §7 --- is that the obstruction to finishing Oltean's proof is a \emph{representation} choice for the language expansion, not the Henkin/Mishra idea itself; replacing the list-substitution remapping with a plain structural map over the syntax tree makes the homomorphism lemmas immediate and lets most of Oltean's scaffolding go through. This is the thread the rest of the paper follows.

A third realization, which became sharp only once the encoding was fixed and the rest of the development compiled, is that the proof invokes freshness in \textbf{two structurally different places}, and Mishra's reservation principle is decisive for one of them and simply inapplicable to the other. At the \textbf{root}, the extended Lindenbaum lemma must witness an \emph{infinite} consistent set, and there Mishra's structural reserve --- Oltean's \texttt{odd\_noms}, the \texttt{N\ ⊕\ ℕ} split internalized in ℕ --- is exactly the right and decisive tool; that part is complete. But the \textbf{truth lemma's ◇-case} must, for each \texttt{◇ψ\ ∈\ Δ}, produce a \emph{witnessed} successor MCS containing \texttt{ψ} together with the box-reduct \texttt{\{χ\ │\ □χ\ ∈\ Δ\}}, and here reservation does not help --- for a reason that has nothing to do with the size of the name supply. For \emph{every} nominal \texttt{j} whatsoever, \texttt{nom\ j\ ⟶\ nom\ j} is a tautology, so \texttt{□(nom\ j\ ⟶\ nom\ j)} is a theorem and lies in every MCS \texttt{Δ}; hence the box-reduct already mentions \emph{all} nominals, reserved ones included. No structural reserve can make a name fresh for that set. The shortcut that tried to force the successor through the same reserve-based Lindenbaum machinery (the lemma \texttt{enough\_noms\_diamond\_seed}) is therefore not merely unproved but \textbf{false}.

The remedy for the successor step is \textbf{not} Mishra's, and it is precisely the direction \textbf{Oltean had already taken}: build the successor by an \emph{existence lemma} in the classical Henkin style, drawing each witness from \texttt{Δ}'s \emph{own} witnessedness through a fresh \emph{state variable} (\texttt{new\_var}) rather than a reserved nominal --- the (already proven) \texttt{l313}/\texttt{l313\textquotesingle{}} lemmas. Oltean's \texttt{set\_family} / \texttt{succesor\_set} scaffolding for this was left incomplete (as \texttt{admit}s), but the \emph{approach} was correct; what remained was to finish it, not to find more fresh names. So the honest division of credit is this: \textbf{Mishra's reservation idea is the right and decisive tool for the root Lindenbaum construction, while Oltean's existence-lemma construction is the right tool for the witnessed successor --- and the work that remained was to complete Oltean's construction, not to extend Mishra's to a place it does not reach.} That construction is now complete (§TL-fix): the accumulating witness family is re-typed to carry \emph{data}, a finite-bounding (compactness) argument over \texttt{diamond\_extension\_consistent} proves the seed consistent, and \texttt{RegularLindenbaumLemma} delivers the witnessed successor MCS. With this last step in place \textbf{the entire completeness theorem is formalized with no remaining \texttt{sorry}/\texttt{admit}.}

\hypertarget{contribution}{%
\subsection{Contribution}\label{contribution}}

This paper:

\begin{enumerate}
\def\labelenumi{\arabic{enumi}.}
\item
  \textbf{Ports} Oltean's development to a current toolchain (Lean v4.30.0 / mathlib v4.30.0), absorbing roughly two and a half years of mathlib API change.
\item
  \textbf{Closes the completeness gap}, completing the witnessed extended Lindenbaum lemma, the existence lemma for completed models, and the final completeness theorem.
\item
  \textbf{Clarifies the design space} for the freshness mechanism that the completeness proof hinges on, and documents the encoding choice that makes the formal proofs go through cleanly.
\end{enumerate}

\hypertarget{the-proof-blueprint-and-the-incoming-state-of-olteans-development}{%
\subsection{The proof blueprint and the incoming state of Oltean's development}\label{the-proof-blueprint-and-the-incoming-state-of-olteans-development}}

Because we are \emph{renovating} an existing, archived formalization rather than writing one from scratch, it is worth stating plainly what we inherited, what already works, and where the genuine difficulty sits --- so that the reader can follow the order in which we attack the problem and understand why some \texttt{admit}s are dispatched in a line while others force a redesign.

\textbf{The blueprint.} Completeness for \emph{L(∀)} is the standard Henkin/canonical-model argument {[}BRV01{]} as adapted to hybrid logic by Blackburn (1998) {[}Bla98{]}, and Oltean's development {[}Olt23{]} wires it up faithfully:

\begin{lstlisting}[style=leanbox]
Γ ⊨ φ  ⟹  Γ ⊢ φ
  └─ via contraposition + Model Existence:  every consistent set is satisfiable
       1. Lindenbaum:           consistent Γ  ⟶  maximal consistent Γ'      [compiles]
       2. Language extension:   reserve an infinite supply of fresh nominals
            ├─ odd_noms:   map the language into the ODD nominals (i ↦ 2i+1)
            └─ pf_extended: ⊢ φ ↔ ⊢ φ⁺   (derivations survive the extension)
       3. Witnessed Lindenbaum: an MCS that witnesses every ◇ / ∃
       4. Completed model + Truth lemma
       5. Existence lemma:      ◇-witnesses provide successor states
       6. assemble  ⟶  Completeness
\end{lstlisting}

This skeleton is sound; the question was never whether the mathematics works (Blackburn proved it on paper) but whether each step survives mechanization in dependent type theory. Soundness, the syntax and substitution machinery, the Kripke semantics, the Hilbert proof system, and ordinary (non-witnessed) Lindenbaum all elaborate and compile.

The dependencies between the remaining deliverables are \textbf{not linear but a directed acyclic graph} (Figure 1): several independent foundations converge on the witnessed Lindenbaum lemma \textbf{G} and again on the final theorem \textbf{I}. Following the now-common practice of stating a Lean development as an explicit blueprint, we record that graph here; nodes are the deliverables of §1.6 (with the already-compiling pieces shaded), and an edge \texttt{X\ →\ Y} means \emph{Y uses X}.

\begin{center}
\includegraphics[max width=\linewidth,max totalheight=0.85\textheight,keepaspectratio]{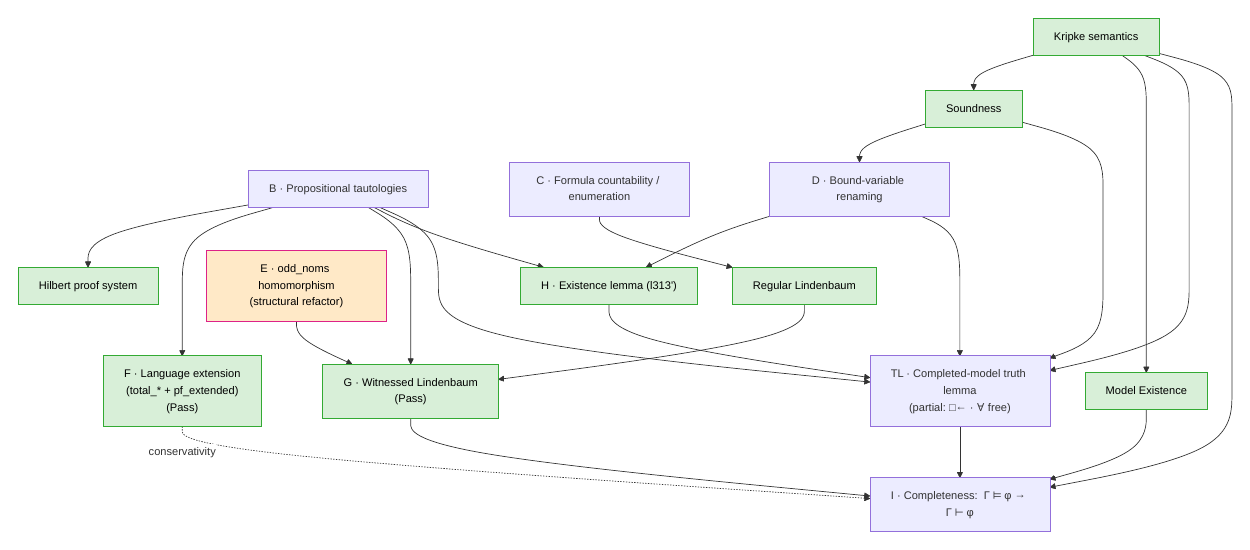}
\end{center}

\textbf{Legend (node colors).} Figure 1 uses three node styles on the \emph{foundation} nodes:

\begin{itemize}
\item
  \textbf{Green} --- pre-existing foundations that already compiled before this work and are \emph{not} deliverables of the completeness effort: Kripke semantics, the Hilbert proof system, Soundness, Regular Lindenbaum, and Model Existence. \textbf{G} is also green now that witnessed Lindenbaum is closed.
\item
  \textbf{Orange} --- the single encoding \emph{crux}, \textbf{E} (\texttt{odd\_noms} homomorphism), discharged by reorganizing the representation rather than by proving the inherited \texttt{admit}s as stated (§1.3).
\item
  \textbf{Blue} --- the deliverables this work closes or is still closing: \textbf{B, C, D, F, TL, H, I} ( \textbf{G} was blue while open; see above).
\end{itemize}

The \textbf{TL} and \textbf{I} subdiagrams (Figures 1a--1d) add \textbf{yellow} = partial / wired but blocked on upstream admits, and \textbf{red} = open \texttt{sorry}/\texttt{admit} rows.

The shading is a snapshot of the \emph{incoming} state; live, per-deliverable status is tracked in the results table (§9).

\emph{Figure 1. Dependency blueprint \ldots{} The two fan-in points, \textbf{G} (now closed) and \textbf{I}, are why the work is a tree rather than a chain.}

\textbf{Module-level snapshots.} Figure 1 is deliberately coarse. Four load-bearing modules each have their own internal order; the diagrams below are sized to fit a single column and are meant to be read \emph{inside} the corresponding deliverable.

\emph{F · language extension (\texttt{LanguageExtension.lean}).} Structural \texttt{total\_*} lemmas are largely independent of \textbf{G}; \textbf{\texttt{pf\_extended} ←} (conservativity) is what unlocks \texttt{consistent\_total} in \textbf{I}, not \texttt{ExtendedLindenbaumLemma}. The backward direction is \textbf{not} a structural induction on \texttt{Proof} (aliens may appear only in subformulas); it follows Blackburn: finitely many alien nominals in \texttt{proof\_noms} → global rename via \texttt{rename\_constants\_fwd} / \texttt{eliminate\_aliens} (F2) → pull back in-range proofs with \texttt{inv\_t} (F3). F1 supplies the \texttt{ax\_q2\_nom} reconstruction lemmas used inside F3.

\emph{Figure 1a · F · language extension.}

\begin{center}
\includegraphics[max width=\linewidth,max totalheight=0.85\textheight,keepaspectratio]{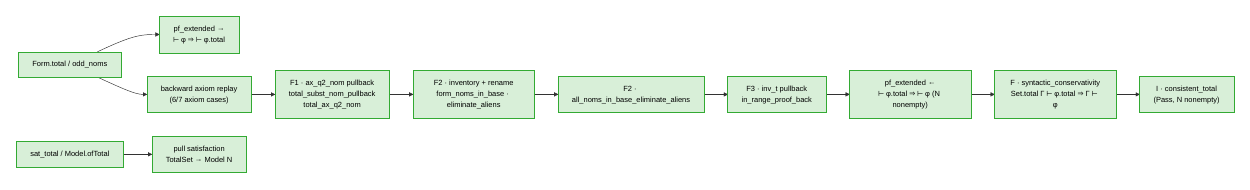}
\end{center}

\emph{Figure 1b · G · witnessed Lindenbaum.} After \textbf{E} makes \texttt{odd\_noms} structural, \textbf{G} is a finiteness argument: each stage adds only finitely many formulas, so some even nominal remains fresh.

\begin{center}
\includegraphics[max width=\linewidth,max totalheight=0.85\textheight,keepaspectratio]{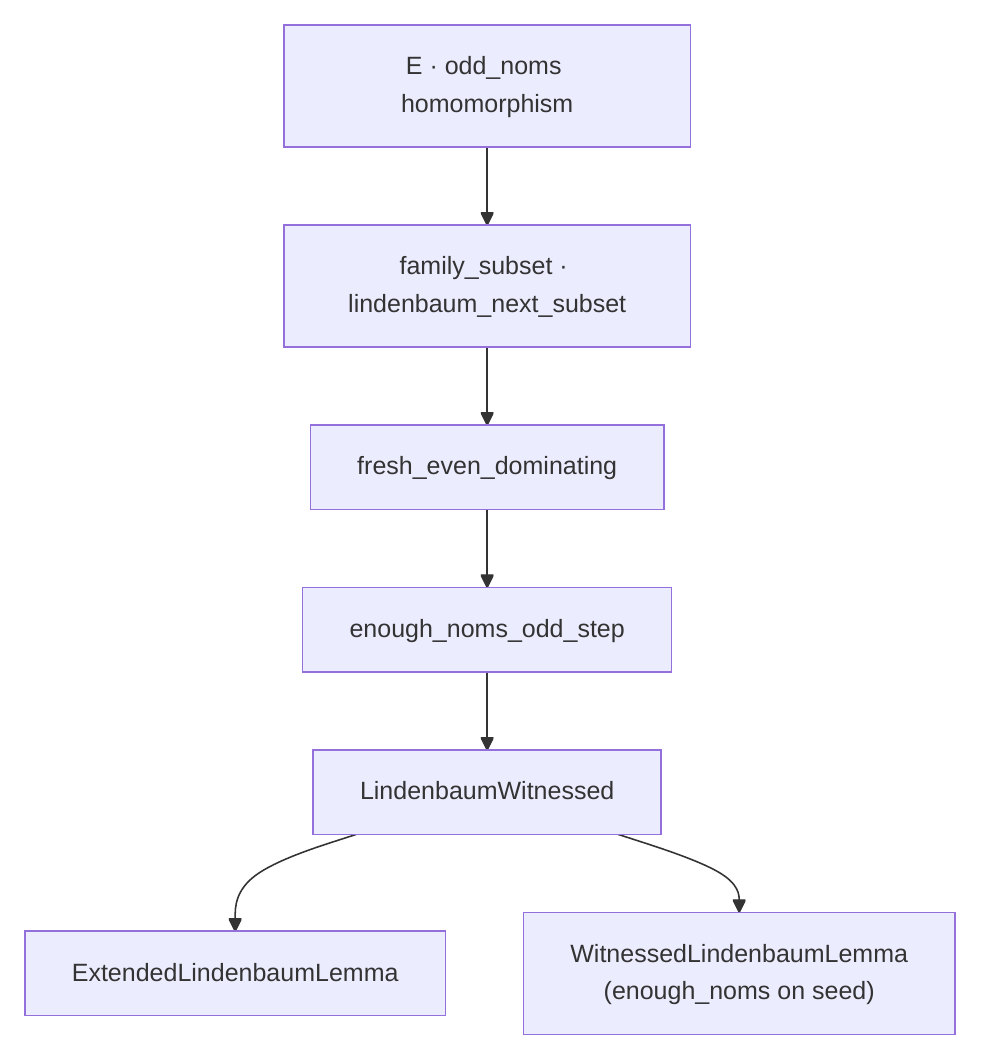}
\end{center}

\emph{\texttt{WitnessedLindenbaumLemma}} (not \texttt{ExtendedLindenbaumLemma}) is what the \textbf{TL} diamond chain calls on the successor seed \texttt{\{ψ\}\ ∪\ \{□χ\ ∈\ Δ\}}.

\emph{Figure 1c · TL · completed-model truth lemma.} Oltean's base cases and \texttt{truth\_ex} compile; \textbf{□} and \textbf{∀} are new. The \textbf{∀} case (\texttt{truth\_all}) is now \textbf{fully closed} for both free and non-free \texttt{x} (uniform proof, dual to \texttt{truth\_ex}); the \textbf{□ →} direction is closed and \textbf{□ ←} runs through the diamond-successor pipeline below (the witnessed ◇-successor existence lemma is now \textbf{complete} via the §TL-fix Henkin construction; the false \texttt{enough\_noms\_diamond\_seed} shortcut has been deleted). \textbf{TruthLemma} is assembled by well-founded recursion on \texttt{Form.depth}, which supplies \texttt{truth\_all}'s depth-indexed induction hypothesis. The whole truth lemma is now \texttt{sorry}-free.

\begin{center}
\includegraphics[max width=\linewidth,max totalheight=0.85\textheight,keepaspectratio]{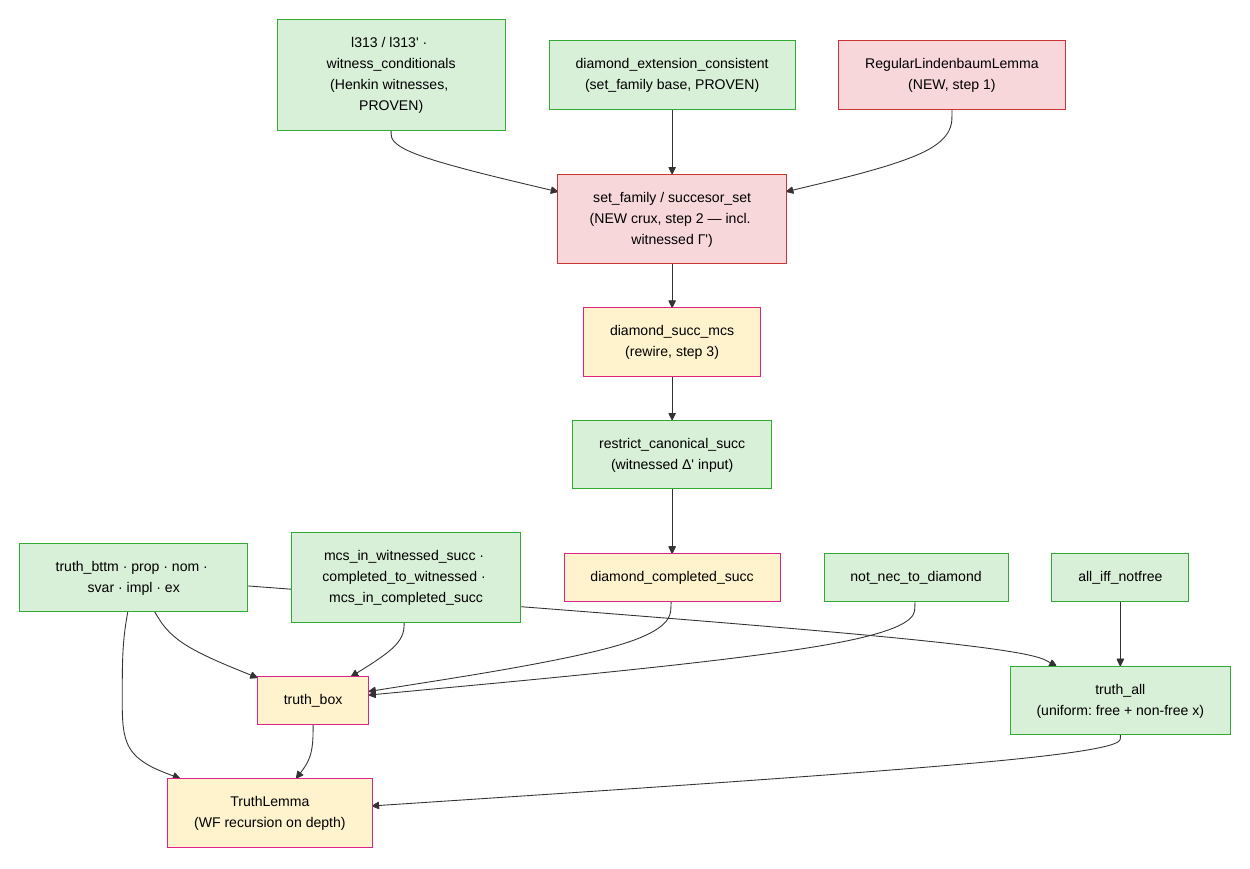}
\end{center}

\emph{Figure 1d · I · model existence.}

\emph{I · model existence (\texttt{Completeness.lean}).} \texttt{cons\_sat} is fully wired; execution still needs backward conservativity and the remaining TL rows below.

\begin{center}
\includegraphics[max width=\linewidth,max totalheight=0.85\textheight,keepaspectratio]{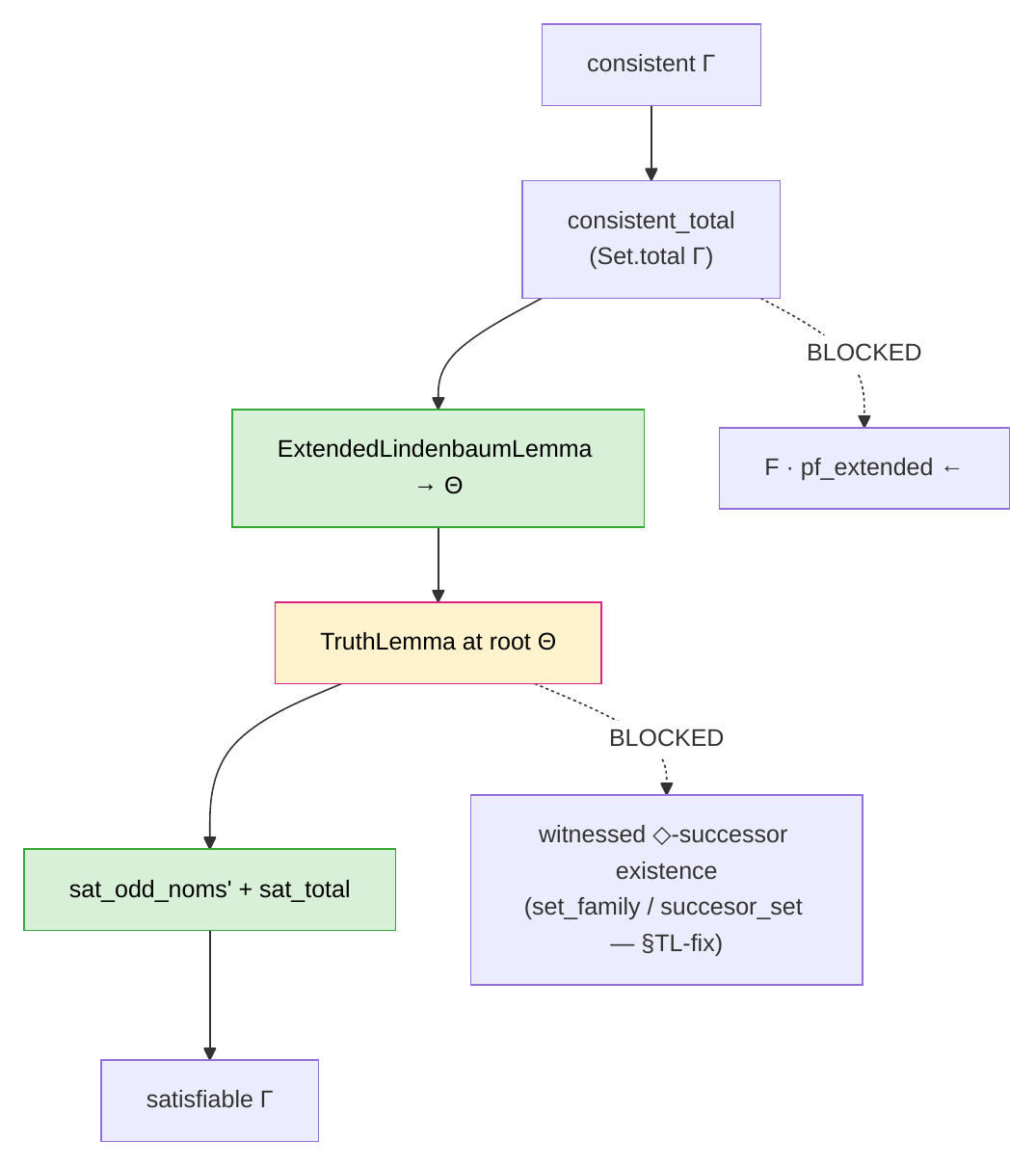}
\end{center}

\textbf{The incoming state: where the holes are.} What Oltean left open is concentrated in the freshness/witnessing layer (steps 2--3) and the pieces that depend on it (the completed model's truth lemma, the existence lemma, and the final assembly). Concretely, the inherited \texttt{sorry}/\texttt{admit} obligations fall into three quite different kinds, and conflating them is what makes ``there are a lot of holes'' sound more alarming than it is:

\begin{enumerate}
\def\labelenumi{\arabic{enumi}.}
\item
  \emph{Mechanical / decidable holes} --- not real mathematical content. The thirteen \texttt{Tautology.lean} truth-table lemmas (one decision-procedure pattern), the formula-countability encoding lemmas, the bound-variable renaming lemmas, and the \texttt{LanguageExtension.total\_*} structural inductions. Also in this bucket, though not \texttt{admit}s but \emph{broken proofs}, is the entire \texttt{CompletedModel} truth lemma: Oltean's proofs there are correct and merely need to be re-fitted to the current \texttt{simp} normal forms in order to compile. These genuinely yield to incremental, local work.
\item
  \emph{Load-bearing, but standard} --- the real Henkin content: witnessed Lindenbaum, the existence lemma, and the final assembly. These are not hard \emph{ideas}; they go through once the layer beneath them is clean.
\item
  \emph{Load-bearing, and an encoding trap} --- the \texttt{odd\_noms} freshness homomorphism (step 2). This is the one place where eliminating the \texttt{admit}s \emph{as stated} is the wrong move.
\end{enumerate}

\textbf{Why one cluster is a trap, and what we do about it.} As explained in §1.3, Oltean realizes structural freshness with \texttt{odd\_noms}, which maps every nominal \texttt{i\ ↦\ 2·i+1} (so the odd nominals carry the image and the even nominals are reserved as a fresh supply --- Mishra's \texttt{N\ ⊕\ ℕ} internalized in \texttt{ℕ}). The \emph{idea} is right. But the \emph{implementation} computes \texttt{odd\_noms\ φ} by collecting φ's nominals into a \textbf{merged, sorted, de-duplicated} list and \texttt{bulk\_subst}-ing along it. Against that representation the apparently trivial homomorphism \texttt{(φ\ ⟶\ ψ).odd\_noms\ =\ φ.odd\_noms\ ⟶\ ψ.odd\_noms} is a real fight with list ordering, deduplication, and no-op substitutions --- and it is precisely this lemma (and its siblings \texttt{odd\_box}, \texttt{odd\_bind}, \texttt{odd\_conj}) that the witnessed Lindenbaum lemma waits on. Discharging these \texttt{admit}s in place would mean proving hard statements about an awkward encoding. The productive move is instead to \textbf{reorganize}: redefine \texttt{odd\_noms} as a plain structural recursion over the syntax tree (\texttt{(φ\ ⟶\ ψ).odd\_noms\ :=\ φ.odd\_noms\ ⟶\ ψ.odd\_noms}, etc.), after which the homomorphism lemmas hold \emph{by definition} (\texttt{rfl}), the freshness property (``no even nominal occurs in \texttt{odd\_noms\ φ}'') becomes a one-line induction, and the supporting \texttt{descending} / \texttt{nocc\_bulk\_property} apparatus is no longer needed. This is the sense in which finishing the proof is partly an exercise in \emph{renovation}: the obstruction is a representation choice, not the construction, and the right response is to change the representation rather than to grind against it.

\textbf{Plan of attack.} We work in the topological order of the blueprint (Figure 1), and where a stage offers a choice we take the \emph{easiest task first}. Concretely: restore the compile (A) so the whole library elaborates with holes marked; clear the decidable/mechanical leaves --- propositional tautologies (B), formula-countability (C), bound-variable renaming (D); carry out the \texttt{odd\_noms} reorganization (E), the one foundation that is a redesign rather than a proof; discharge the language-extension structural lemmas (\textbf{F}, the \texttt{total\_*} batch --- largely parallel to \textbf{E}); prove the witnessed Lindenbaum lemma (\textbf{G}), which in the code depends chiefly on \textbf{C}, \textbf{E}, and \textbf{B}; close the existence lemma (\textbf{H}), which depends on \textbf{B} and \textbf{D} only; discharge the language-extension structural lemmas (\textbf{F}, the \texttt{total\_*} batch --- parallel to \textbf{E}); finish \textbf{F}'s conservativity half (\texttt{pf\_extended} ←), which feeds \textbf{I} but not \textbf{G} or \textbf{H}; re-fit the completed-model truth lemma (\textbf{TL}), which waits on \textbf{H}; and assemble the final theorem (\textbf{I}). In §9, \textbf{G} and \textbf{H} are listed before \textbf{F} so \textbf{Pass} rows are not buried under \textbf{F}'s open conservativity substeps.

\hypertarget{goal-and-major-steps}{%
\subsection{Goal and major steps}\label{goal-and-major-steps}}

\textbf{Goal.} Produce a fully \texttt{sorry}-free Lean 4 proof (under Lean v4.30.0 / mathlib v4.30.0) of the \textbf{completeness theorem} for \emph{L(∀)} --- \texttt{(Γ\ ⊨\ φ)\ →\ (Γ\ ⊢\ φ)} --- finishing the construction Oltean left open. Soundness, syntax, semantics, and most scaffolding already exist; the gap is the Henkin-style completeness argument and the ``freshness'' machinery it depends on.

The work decomposes into the following major steps. Letter labels \textbf{A}--\textbf{I} follow Oltean's proof narrative; §9 lists \textbf{G} and \textbf{H} before \textbf{F} because those steps are \textbf{Pass} in the code and do not import \texttt{LanguageExtension} or wait on \texttt{pf\_extended} ← (only \textbf{I} does). Status is tracked in the results table (§9). of \texttt{sorry}/\texttt{admit} obligations \emph{inherited from Oltean's development}; we group them by the mathematical reason they exist. (We verified against the archived upstream sources that these holes are Oltean's own, not artifacts of our port --- for instance Oltean's original \texttt{Tautology.lean} already carries the thirteen \texttt{admit}s below.)

\begin{itemize}
\item
  \textbf{A. Get the whole library compiling.} Fix roughly two and a half years of mathlib API churn module-by-module in dependency order so that \texttt{lake\ build} succeeds with the proof holes still marked \texttt{sorry}/\texttt{admit}. (Per-module status is tracked in §9; the larger re-fit of \texttt{CompletedModel}'s truth lemma is split out as its own step, \textbf{TL}, since it feeds the final assembly \textbf{I} in the blueprint.)
\item
  \textbf{B. Remove the propositional-tautology holes.} Discharge the \texttt{Tautology.lean} truth-table lemmas Oltean left as \texttt{admit} (\texttt{hs\_taut}, \texttt{neg\_intro}, \texttt{conj\_intro}, \texttt{conj\_intro\_hs}, \texttt{iff\_intro}, \texttt{iff\_elim\_l}, \texttt{iff\_elim\_r}, \texttt{iff\_rw}, \texttt{iff\_imp}, \texttt{disj\_intro\_l}, \texttt{disj\_intro\_r}, \texttt{disj\_elim}, \texttt{mp\_help}) plus \texttt{ProofUtils.iff\_subst}. All are decidable propositional facts.
\item
  \textbf{C. Remove the formula-countability holes.} \texttt{FormCountable}: \texttt{prime\_2\_3} (a number-theoretic fact, \texttt{3\^{}(n+1)\ ≠\ 2\^{}(m+1)}), \texttt{guns}, and \texttt{of\_brixton} --- injectivity bookkeeping for the Gödel-style encoding that makes \texttt{Form} countable (needed to enumerate formulas for Lindenbaum).
\item
  \textbf{D. Remove the bound-variable-renaming holes.} \texttt{RenameBound}: \texttt{replace\_neg}, \texttt{replace\_bound\_depth}, and \texttt{substable\_after\_replace} --- structural facts about α-renaming bound state variables.
\item
  \textbf{E. Remove the structural-freshness homomorphism holes (the crux).} \texttt{Substitutions}: \texttt{bulk\_subst\_impl}, \texttt{list\_noms\_impl\_r}, \texttt{list\_noms\_impl\_l}, \texttt{odd\_box}, \texttt{odd\_bind}, \texttt{List.to\_odd}, \texttt{List.odd\_to}, \texttt{odd\_conj}, \texttt{odd\_conj\_rev} --- that Oltean's \texttt{i\ ↦\ 2·i+1} remapping (\texttt{odd\_noms}) is a homomorphism for the connectives and conjunctions. This is where Oltean's \texttt{bulk\_subst}-over-sorted-lists encoding makes the ``obvious'' lemmas hard (§1.3), and everything downstream depends on it.
\item
  \textbf{G. Remove the witnessed-Lindenbaum holes.} \texttt{Lindenbaum}: \texttt{LindenbaumWitnessed} and \texttt{ExtendedLindenbaumLemma}. In the Lean graph this module imports \textbf{E} / countability / proof scaffolding only --- not \texttt{LanguageExtension}.
\item
  \textbf{H. Remove the existence-lemma hole.} \texttt{ExistenceLemma.l313\textquotesingle{}}: the diamond-witness property used to build successor states of the completed model. Depends on \textbf{B} and \textbf{D} only (Figure 1); does not use \textbf{F} or \textbf{G} (\texttt{l313\textquotesingle{}} is on base-language \texttt{Form\ N}, not \texttt{TotalSet} / \texttt{pf\_extended}).
\item
  \textbf{F. Remove the language-extension / theorem-preservation holes.} \texttt{LanguageExtension}: structural \texttt{total\_*} lemmas, \texttt{l416}, and \texttt{pf\_extended} (Prop. 4.1.7: derivations survive the language expansion). The \textbf{\texttt{total\_*} block is largely independent of }G** and \textbf{H}. \textbf{\texttt{pf\_extended} ←} (conservativity: F1 \texttt{ax\_q2\_nom} pullback, F2 alien elimination, F3 \texttt{inv\_t} pullback) is now \textbf{complete}, together with \texttt{syntactic\_conservativity} (the \texttt{Set.total\ Γ\ ⊢\ φ.total\ ⇒\ Γ\ ⊢\ φ} lift). This is load-bearing for \textbf{I} (\texttt{consistent\_total}), not for \texttt{ExtendedLindenbaumLemma} or \texttt{l313\textquotesingle{}}. This path is now \textbf{complete}: \texttt{consistent\_total} is proven and the \texttt{N}-nonempty hypothesis (needed to pick a base nominal for alien elimination) is threaded through \texttt{cons\_sat} / \texttt{Completeness}. The former last obstacle --- the \textbf{TL} witnessed ◇-successor existence lemma --- is now discharged by the §TL-fix Henkin construction (\texttt{enough\_noms\_diamond\_seed} was false and has been deleted), so the development is complete.
\item
  \textbf{TL. Re-fit the completed-model truth lemma.} \texttt{CompletedModel}: restore Oltean's truth-lemma cases (\texttt{truth\_bttm}, \texttt{truth\_prop}, \texttt{truth\_nom}, \texttt{truth\_svar}, \texttt{truth\_impl}, \texttt{truth\_ex}) and the supporting valuation lemmas to the current \texttt{simp} normal forms. \textbf{\texttt{truth\_box} and \texttt{truth\_all} are new} --- Oltean's archived development stops before the modal/binder cases. \texttt{TruthLemma} is assembled by well-founded recursion on \texttt{Form.depth}; the \texttt{bind} case delegates to \texttt{truth\_all}, now \textbf{fully closed} for both free and non-free \texttt{x} (uniform \texttt{has\_state\_symbol} split + depth-indexed \texttt{ih}, dual to \texttt{truth\_ex}). The \textbf{□ ←} witnessed ◇-successor existence lemma --- the former last obstacle --- is now closed by the \texttt{l313\textquotesingle{}}-based Henkin construction (\texttt{succ\_seed} + \texttt{RegularLindenbaumLemma}); the false \texttt{enough\_noms\_diamond\_seed} shortcut has been deleted. See \textbf{§TL-fix} for the disproof and the completed construction. Depends on \textbf{B}, \textbf{D}, \textbf{H} (and on Kripke semantics and Soundness).
\item
  \textbf{I. Remove the final-completeness hole.} \texttt{Completeness}: \texttt{cons\_sat} runs \texttt{consistent\_total} → \texttt{ExtendedLindenbaumLemma\ (Set.total\ Γ)} → \texttt{TruthLemma} at the root witnessed MCS → \texttt{sat\_odd\_noms\textquotesingle{}} / \texttt{sat\_total}; \texttt{Completeness} is then \texttt{ModelExistence} + contraposition. \textbf{\texttt{pf\_extended} forward is not on this path}; only backward conservativity feeds \texttt{consistent\_total}.
\end{itemize}

The substantive mathematics is concentrated in \textbf{E}--\textbf{I}; \textbf{B}--\textbf{D} are essentially mechanical leaf lemmas. \textbf{E} is the crux, for the encoding reasons discussed in §1.3.

\hypertarget{tl-fix-the-witnessed--successor-existence-lemma-resolved}{%
\subsection{§TL-fix · The witnessed ◇-successor existence lemma (resolved)}\label{tl-fix-the-witnessed--successor-existence-lemma-resolved}}

\begin{quote}
\textbf{Status: complete.} The construction below is fully formalized; \texttt{enough\_noms\_diamond\_seed} has been deleted and \texttt{diamond\_succ\_mcs} is rewired onto it. \texttt{\#print\ axioms\ Completeness} reports only \texttt{propext,\ Classical.choice,\ Quot.sound}. The step plan is retained as the record of how the last obstacle was discharged.
\end{quote}

\textbf{Why \texttt{enough\_noms\_diamond\_seed} is false (not just hard).} The lemma claims \texttt{enough\_noms\ (\{ψ\}\ ∪\ \{χ\ │\ □χ\ ∈\ Δ\})}, whose first conjunct (\texttt{enough\_noms}, \texttt{Lindenbaum.lean}) demands a nominal \texttt{i} occurring in \textbf{no} formula of the set. But for \emph{every} nominal \texttt{i}, \texttt{nom\ i\ ⟶\ nom\ i} is a tautology, so \texttt{⊢\ □(nom\ i\ ⟶\ nom\ i)} by necessitation, so \texttt{□(nom\ i\ ⟶\ nom\ i)\ ∈\ Δ} for any MCS \texttt{Δ}; hence \texttt{(nom\ i\ ⟶\ nom\ i)\ ∈\ \{χ\ │\ □χ\ ∈\ Δ\}} and \texttt{nom\_occurs\ i\ (nom\ i\ ⟶\ nom\ i)\ =\ true}. So \texttt{all\_nocc\ i} fails for \emph{every} \texttt{i}: the box-reduct of any MCS mentions all nominals, and there is no reserve to be had --- independent of how \texttt{Δ} was built. The \texttt{WitnessedLindenbaumLemma}-on-the-seed approach is therefore structurally unworkable; it requires a globally fresh nominal that provably does not exist.

\textbf{The correct route (Oltean's intended Henkin construction).} Build the witnessed successor \emph{incrementally}, borrowing witnesses from \texttt{Δ}'s own witnessedness via \texttt{l313\textquotesingle{}} --- which uses a fresh \textbf{variable} (\texttt{new\_var}), not a fresh nominal. The hardest analytic lemma (\texttt{l313}/\texttt{l313\textquotesingle{}}) and the witness-conditional accumulator are \textbf{already proven} (\texttt{ExistenceLemma.lean}, live code), and \textbf{\texttt{RegularLindenbaumLemma} already exists} (\texttt{Lindenbaum.lean}, general over any \texttt{N}). A reconnaissance pass turned up the precise obstruction and a concrete plan:

\textbf{The data-vs-\texttt{Prop} flaw.} \texttt{witness\_conditionals} currently returns \texttt{∃\ l,\ l\ ≠\ {[}{]}\ ∧\ ◇conjunction\textquotesingle{}\ l\ ∈\ Δ} --- a \textbf{\texttt{Prop}} --- and \texttt{succesor\_set}/\texttt{succesor\_set\textquotesingle{}} extract the list with \texttt{.choose}. Because \texttt{Exists} is proof-irrelevant, \texttt{.choose} returns \emph{some} list with that property, \textbf{not} the structured accumulating one the recursion built; the ``the witness conditional for \texttt{enum\ n} is in the list'' fact is then unrecoverable, and witnessedness cannot be proven. \emph{This is exactly why the commented \texttt{set\_family}/\texttt{succesor\_set} stalled.} The fix is to return \textbf{data} (a \texttt{Subtype}/\texttt{Sigma}), preserving the list.

\textbf{How it was discharged (all steps done):}

\begin{itemize}
\item
  \textbf{2.0 ✓} Re-typed the accumulator (\texttt{wcond} / \texttt{wcond\_step}, \texttt{ExistenceLemma.lean}) to \texttt{\{\ l\ :\ List\ (Form\ N)\ //\ l\ ≠\ {[}{]}\ ∧\ ◇conjunction\textquotesingle{}\ l\ ∈\ Δ\ \}}, preserving the recursion (\texttt{{[}ψ{]}} at the base, prepend \texttt{((ex\ x,σ)⟶σ{[}i//x{]})} from \texttt{l313\textquotesingle{}} at each existential step). The index \texttt{i} is \texttt{l313\textquotesingle{}}'s \texttt{.choose}; crucially the \emph{list} is now data, so its members are recoverable.
\item
  \textbf{2.1 ✓} \texttt{wcond\_succ\_mem} / \texttt{wcond\_mono} (stage membership is monotone in the index) and \texttt{wcond\_step\_mem} (\texttt{enum\ n\ =\ ex\ x,σ\ →\ ∃\ i,\ ((ex\ x,σ)⟶σ{[}i//x{]})\ ∈\ (wcond\ (n+1)).val}, proved by iota-reducing \texttt{wcond\_step} on the literal \texttt{ex\ x,σ}).
\item
  \textbf{2.2 ✓} \texttt{succ\_seed\ :=\ \{χ\ │\ □χ\ ∈\ Δ\}\ ∪\ \{χ\ │\ ∃\ n,\ χ\ ∈\ (wcond\ n).val\}} (so \texttt{ψ\ ∈\ succ\_seed} at stage 0, and \texttt{\{χ│□χ∈Δ\}\ ⊆\ succ\_seed}).
\item
  \textbf{2.3 ✓} \texttt{succ\_seed\_consistent} (\texttt{CompletedModel.lean}): \texttt{seed\_list\_bound} puts any finite \texttt{L\ ⊆\ succ\_seed} inside the box-reduct together with a single stage \texttt{wcond\ N} (\texttt{wcond\_mono}); then \texttt{box-reduct\ ∪\ \{conjunction\textquotesingle{}\ (wcond\ N).val\}} derives \texttt{conjunction\ succ\_seed\ L} (\texttt{conj\textquotesingle{}\_imp\_mem} for the conditionals, \texttt{Γ\_premise} for the box part), and since \texttt{◇conjunction\textquotesingle{}\ (wcond\ N).val\ ∈\ Δ}, \texttt{diamond\_extension\_consistent} (applied to that conjunction) closes it --- the same \texttt{box\_of\_consequence}/\texttt{MCS\_mp} finish as the base case.
\item
  \textbf{2.4 ✓} \texttt{RegularLindenbaumLemma\ succ\_seed} → MCS \texttt{Γ\textquotesingle{}\ ⊇\ succ\_seed}.
\item
  \textbf{2.5 ✓} Output properties: \textbf{\texttt{Canonical.R\ Δ\ Γ\textquotesingle{}}} (box-reduct ⊆ \texttt{Γ\textquotesingle{}}); \textbf{\texttt{ψ\ ∈\ Γ\textquotesingle{}}} (stage 0); \textbf{\texttt{witnessed\ Γ\textquotesingle{}}} --- \texttt{enum\ =\ f.invFun} is surjective (left inverse of the injection from \texttt{exists\_injective\_nat}), so any \texttt{ex\ x,σ\ ∈\ Γ\textquotesingle{}} is \texttt{enum\ (f\ (ex\ x,σ))}; its conditional is in \texttt{wcond\ (·+1)\ ⊆\ succ\_seed\ ⊆\ Γ\textquotesingle{}}, and \texttt{MCS\_mp} yields \texttt{σ{[}i//x{]}\ ∈\ Γ\textquotesingle{}}. \emph{(This is the milestone Oltean stalled on; with the data refactor it reduces to \texttt{MCS\_mp} + surjectivity.)}
\item
  \textbf{2.6 ✓ (Step 3)} \texttt{diamond\_succ\_mcs} now returns \texttt{⟨Γ\textquotesingle{},\ Canonical.R\ Δ\ Γ\textquotesingle{},\ ψ∈Γ\textquotesingle{},\ MCS\ Γ\textquotesingle{},\ \ \ witnessed\ Γ\textquotesingle{}⟩} from this construction; \texttt{enough\_noms\_diamond\_seed} is \textbf{deleted} (\texttt{diamond\_extension\_consistent} is retained --- it powers 2.3).
\end{itemize}

Steps 2.0--2.6 turned the TL \texttt{Partial} rows and the two I \texttt{Partial} rows (\texttt{cons\_sat}, \texttt{Completeness}) green, finishing the whole development. The decisive new technical content was the data refactor (2.0--2.1) and the compactness bookkeeping (2.3); no fundamental wall remained (the box-leak that kills \texttt{enough\_noms} does not affect this route).

\emph{Attribution (cf.~§1.3).} This step is \textbf{not} an application of Mishra's structural-freshness suggestion --- that idea is decisive at the \emph{root} Lindenbaum construction but inapplicable here, since the box-reduct \texttt{\{χ\ │\ □χ\ ∈\ Δ\}} mentions every nominal (\texttt{□(nom\ j\ ⟶\ nom\ j)\ ∈\ Δ} for all \texttt{j}). The witnessed successor is instead built by \textbf{Oltean's existence-lemma direction} (\texttt{l313\textquotesingle{}}, fresh \emph{variable} + \texttt{Δ}'s witnessedness), which was correct but left incomplete; the work here was to finish it --- now done.

\begin{center}\rule{0.5\linewidth}{0.5pt}\end{center}

\hypertarget{background-the-logic-l}{%
\section{\texorpdfstring{Background: the logic \emph{L(∀)}}{Background: the logic L(∀)}}\label{background-the-logic-l}}

\emph{(Condensed; full definitions follow Blackburn 1998 and Oltean's thesis.)}

\textbf{Signature.} A hybrid signature is a triple ⟨PROP, SVAR, NOM⟩ of denumerable sets of propositional symbols, state variables, and nominals.

\textbf{Formulas.} \texttt{φ\ ::=\ ⊥\ \textbar{}\ a\ \textbar{}\ φ\ →\ φ\ \textbar{}\ □φ\ \textbar{}\ ∀x\ φ}, where \texttt{a} ranges over atomic symbols (propositions, state variables, nominals) and \texttt{x} over state variables. Negation, conjunction, ◇, and ∃ are defined as usual.

\textbf{Semantics.} A model \texttt{M\ =\ ⟨W,\ R,\ V⟩} is a Kripke frame with a valuation; an assignment \texttt{g} sends each state variable to a single state. Nominals and state variables denote singletons. Satisfaction \texttt{M,\ s,\ g\ ⊨\ φ} is standard, with \texttt{M,\ s,\ g\ ⊨\ x} iff \texttt{g(x)\ =\ \{s\}} and \texttt{M,\ s,\ g\ ⊨\ ∀x\ φ} iff φ holds at \texttt{s} under every \texttt{x}-variant of \texttt{g}.

\textbf{Proof system.} A Hilbert system with classical tautologies, axiom K, the quantifier axioms (Q1, Q2 for variables and nominals), Name, Nom, Barcan, and the rules modus ponens, generalization, and necessitation. \texttt{Γ\ ⊢\ φ} is syntactic consequence.

\begin{center}\rule{0.5\linewidth}{0.5pt}\end{center}

\hypertarget{completeness-via-witnessed-maximal-consistent-sets}{%
\section{Completeness via witnessed maximal consistent sets}\label{completeness-via-witnessed-maximal-consistent-sets}}

The completeness proof follows the Henkin/canonical-model method as adapted to hybrid logic by Blackburn (1998):

\begin{enumerate}
\def\labelenumi{\arabic{enumi}.}
\item
  \textbf{Lindenbaum's lemma.} Every consistent set extends to a maximal consistent set (MCS).
\item
  \textbf{Witnessed MCSs.} An MCS Δ is \emph{witnessed} if whenever \texttt{∃x\ φ\ ∈\ Δ} there is a nominal \texttt{i} with \texttt{(∃x\ φ\ →\ φ{[}i/x{]})\ ∈\ Δ}. Existence of witnessed MCSs requires \emph{enough nominals}: each existential needs its own witness, and to extend an \emph{infinite} consistent set we need an infinite reserve of nominals that do not already occur.
\item
  \textbf{Language expansion.} To guarantee enough witnesses, expand the language with a denumerable set of new nominals (\texttt{L(∀)\ ⊆\ L⁺(∀)}). Expansion is truth-preserving (semantically obvious) and theorem-preserving (Prop. 4.1.7 in the thesis: a derivation using extra nominals can be replayed with those nominals replaced by fresh variables).
\item
  \textbf{Canonical / completed model.} Build the canonical model from MCSs; restrict to a generated, witnessed submodel so that state symbols name uniquely; ``glue on'' a dummy state only when needed to make the model standard.
\item
  \textbf{Truth lemma + existence lemma}, yielding completeness via the model-existence theorem.
\end{enumerate}

\hypertarget{the-freshness-obstacle}{%
\subsection{The freshness obstacle}\label{the-freshness-obstacle}}

Steps 2--3 are where the formalization stalls. Mathematically one simply says ``let \texttt{i₀,\ i₁,\ …} enumerate the new nominals'' and uses \texttt{iₙ} as the witness at step \texttt{n}. In \textbf{set theory} there is never a shortage of fresh names. In \textbf{dependent type theory}, a type already contains \emph{all} its inhabitants: given \texttt{N\ :\ Type}, there is in general no \texttt{N\textquotesingle{}\ ⊋\ N}. One can dynamically search for an unused nominal of a formula (finitely many occur), but to witness an \emph{infinite} Lindenbaum union one must reserve infinitely many nominals \emph{globally} and prove they never occur --- and doing that bookkeeping by dynamic search is exactly what Oltean found ``prohibitively difficult''.

\begin{center}\rule{0.5\linewidth}{0.5pt}\end{center}

\hypertarget{structural-freshness}{%
\section{Structural freshness}\label{structural-freshness}}

The resolution is to make freshness \textbf{structural} rather than computed: arrange the language so that an infinite family of nominals is, by construction, disjoint from the nominals any formula of the base language can use. Three concrete realizations:

\begin{itemize}
\item
  \textbf{Disjoint sum (Mishra).} Parameterize formulas by a nominal type and extend it to \texttt{N\ ⊕\ ℕ}. Witnesses are drawn exclusively from the right summand \texttt{Sum.inr\ n}, which is \emph{structurally} distinct from every base nominal \texttt{Sum.inl\ \_}. Freshness is then a triviality of the sum type, never a search.
\item
  \textbf{Odd/even split inside ℕ (Oltean).} Take a single nominal type \texttt{TotalSet\ ≅\ ℕ} and remap every nominal \texttt{i\ ↦\ 2·i+1} (\texttt{Form.odd\_noms}). The image uses only \emph{odd} nominals, so \emph{all even} nominals are reserved as a fresh supply. This is the same disjoint-sum idea, internalized in ℕ (odds ≅ \texttt{Sum.inl}, evens ≅ \texttt{Sum.inr}), and is the route taken in the existing development.
\item
  \textbf{Abstract name supply (From).} Work over a fixed type with an infinite set of names plus an abstract \texttt{fresh\ :\ Finset\ Name\ →\ Name} returning an unused name, and factor the witnessing into a reusable, logic-generic Lindenbaum/saturation lemma.
\end{itemize}

These are not competitors at the conceptual level --- all three reserve an infinite disjoint supply of names. They differ in \emph{how the reservation is encoded}, and that choice determines how painful the surrounding lemmas are.

\begin{center}\rule{0.5\linewidth}{0.5pt}\end{center}

\hypertarget{related-work}{%
\section{Related work}\label{related-work}}

\begin{itemize}
\item
  \textbf{Asta Halkjær From} gave the \emph{first} machine-checked completeness proof for any hybrid logic --- a Seligman-style tableau system for \textbf{basic} hybrid logic (nominals and satisfaction operators \texttt{@}, \emph{no binders}) in Isabelle/HOL (TYPES 2020) {[}Fro20{]} --- and later an abstract, transfinite synthetic-completeness \emph{framework} (\emph{An Isabelle/HOL Framework for Synthetic Completeness Proofs}, CPP 2025) {[}Fro25{]} instantiated to propositional, first-order, modal, and (basic) hybrid logic. This is the closest existing mechanization of witnessed/named MCSs for hybrid logic and the state of the art for reusable completeness infrastructure. Our target differs in the \emph{object logic}: \emph{L(∀)} is a \textbf{binding} hybrid logic (the satisfaction-style universal binder ∀), proved complete here via a Hilbert system rather than a tableau or natural-deduction calculus. We make no priority claim over From for hybrid logic in general; we are simply not aware of a prior mechanized completeness proof for a \emph{binding} hybrid logic, or of any hybrid-logic completeness in Lean, and we leave that bookkeeping to the reader.
\item
  Earlier Lean modal-logic formalizations: a Henkin-style completeness proof for \textbf{S5} (Bentzen 2021) {[}Ben21{]}, \textbf{Public Announcement Logic / PAL-S5} (Li 2020) {[}Li20{]}, and \textbf{Applicative Matching Logic} in Lean (Cheval \& Macovei 2023) {[}CM23{]}. We are not aware of a prior completeness formalization for a \emph{binding} hybrid logic in any proof assistant, nor of any prior hybrid-logic completeness formalization in Lean.
\item
  The mathematics followed throughout is \textbf{Blackburn}, \emph{Hybrid Completeness} (1998) {[}Bla98{]}.
\end{itemize}

\begin{center}\rule{0.5\linewidth}{0.5pt}\end{center}

\hypertarget{the-lean-4-development}{%
\section{The Lean 4 development}\label{the-lean-4-development}}

The development is \textbf{17 modules, ≈7,245 lines} of Lean 4, pinned to \textbf{Lean v4.30.0 / mathlib v4.30.0}. The complete source is inlined in Appendix A; per-declaration status is tabulated in §9. In \texttt{Hybrid.lean} import order the modules group into four layers:

\begin{itemize}
\item
  \textbf{Syntax and substitution.} \texttt{Util}, \texttt{Form} (formulas, \texttt{Form.depth}), \texttt{Tautology} (propositional reasoning), and \texttt{Substitutions} (state-variable and nominal substitution \texttt{φ{[}i\ //\ x{]}}) fix the object language.
\item
  \textbf{Proof system and soundness.} \texttt{Proof} (the Hilbert calculus, \texttt{MCS}, \texttt{consistent}, \texttt{witnessed}), \texttt{ProofUtils} and \texttt{ListUtils} (derived rules, \texttt{conjunction}, deduction theorem, \texttt{MCS\_mp}, \texttt{box\_of\_consequence}), \texttt{Truth} (Kripke semantics, \texttt{⊨}), and \texttt{Soundness} (\texttt{⊢\ φ\ →\ ⊨\ φ}).
\item
  \textbf{The freshness machinery.} \texttt{RenameBound} and \texttt{FormCountable} (bound-variable renaming and a Gödel-style enumeration of \texttt{Form}), \texttt{Lindenbaum} (\texttt{RegularLindenbaumLemma}, and the witnessed extension \texttt{LindenbaumWitnessed} / \texttt{ExtendedLindenbaumLemma}), and \texttt{LanguageExtension} --- the structural-freshness layer (\texttt{Form.odd\_noms}, the conservativity certificate \texttt{pf\_extended\ :\ ⊢\ φ\ ↔\ ⊢\ φ⁺}, \texttt{syntactic\_conservativity}, and the alien-elimination route feeding \texttt{consistent\_total}).
\item
  \textbf{Canonical model and completeness.} \texttt{ExistenceLemma} (the witnessed ◇-successor: \texttt{l313\textquotesingle{}}, the data-carrying accumulator \texttt{wcond}, the seed \texttt{succ\_seed}, and its consistency \texttt{succ\_seed\_consistent}), \texttt{CompletedModel} (\texttt{diamond\_succ\_mcs}, and the truth lemma \texttt{TruthLemma} assembled by well-founded recursion on \texttt{Form.depth}, with the universal case \texttt{truth\_all} and the modal case \texttt{truth\_box}), and \texttt{Completeness} (the theorem \texttt{Γ\ ⊨\ φ\ →\ Γ\ ⊢\ φ}, assembled from model existence and contraposition, taking \texttt{N} nonempty for alien elimination).
\end{itemize}

The two load-bearing constructions are exactly the two discussed in §1 and §3--§4. \emph{Structural freshness} (\texttt{LanguageExtension.odd\_noms} + \texttt{ExtendedLindenbaumLemma}) supplies the root witnessed MCS; the \emph{existence-lemma Henkin route} (\texttt{wcond}/\texttt{succ\_seed}/\texttt{succ\_seed\_consistent} → \texttt{diamond\_succ\_mcs}) supplies the witnessed ◇-successor. The development is \textbf{\texttt{sorry}-free}, and \texttt{\#print\ axioms\ Completeness} reports only \texttt{propext}, \texttt{Classical.choice}, and \texttt{Quot.sound}.

\begin{center}\rule{0.5\linewidth}{0.5pt}\end{center}

\hypertarget{discussion-encoding-choices}{%
\section{Discussion: encoding choices}\label{discussion-encoding-choices}}

The recurring lesson of this development is that the \emph{mathematical} ideas were settled --- the difficulty lived almost entirely in \textbf{representation choices}, and the same kind of choice surfaced twice.

\textbf{Freshness: which encoding.} All three realizations of structural freshness in §4 are conceptually equivalent --- reserve an infinite, disjoint supply of names. What differs is cost. The disjoint sum \texttt{N\ ⊕\ ℕ} (Mishra) makes freshness a triviality of the sum type but re-parameterizes every formula by a new nominal type, rippling through the entire inherited development. From's abstract \texttt{fresh\ :\ Finset\ Name\ →\ Name} supply is the most reusable but presupposes a logic-generic saturation framework we did not have in Lean. The odd/even split inside \texttt{ℕ} (\texttt{Form.odd\_noms}, \texttt{i\ ↦\ 2·i+1}) keeps the original nominal type and so disturbs the least; it is the encoding we adopt. Crucially, the obstacle that had stalled the proof was \emph{not} the odd/even idea but its \textbf{representation}: Oltean implemented \texttt{odd\_noms} as an iterated single-nominal \texttt{bulk\_subst} over a merged, sorted, de-duplicated list of a formula's nominals, against which the otherwise-trivial homomorphism \texttt{(φ\ ⟶\ ψ).odd\_noms\ =\ φ.odd\_noms\ ⟶\ ψ.odd\_noms} became a fight with list ordering and deduplication. Replacing the list-substitution remap with a plain structural recursion over the syntax tree makes the homomorphism hold by \texttt{rfl}, the theorem-preservation lemmas (\texttt{pf\_extended}) go through, and most of the surrounding scaffolding disappears. The conclusion is blunt: \emph{finishing the proof was largely a matter of choosing the representation, not grinding against it.}

\textbf{Witnessing: \texttt{Prop} versus data.} The ◇-successor exposed the same theme in a different key. Oltean's existence-lemma scaffolding (\texttt{set\_family}/\texttt{succesor\_set}) had stalled because \texttt{witness\_conditionals} returned an existential in \texttt{Prop}: the witness list, recovered via \texttt{.choose}, was proof-irrelevant, so the structure built by recursion was lost and witnessedness could not be re-derived. Re-typing the accumulator as a \texttt{Subtype} (\texttt{wcond}, carrying the list as \emph{data}) preserves exactly the information the compactness argument in \texttt{succ\_seed\_consistent} needs. No new mathematics --- a \texttt{Prop}-to-data representation change.

\textbf{Porting.} Oltean's development targeted a June-2023 Lean nightly; we pin to Lean v4.30.0 / mathlib v4.30.0. The bulk of the porting effort was mathlib API churn --- shifted \texttt{simp} normal forms, renamed lemmas, and changed implicit-argument counts in the \texttt{List} API --- rather than mathematical change. The single largest item was re-fitting the completed-model truth lemma (\texttt{truth\_bttm}/\texttt{prop}/\texttt{nom}/\texttt{svar}/\texttt{impl}/\texttt{ex} and the new \texttt{truth\_all}/\texttt{truth\_box}) to current \texttt{simp} normal forms; this is the kind of maintenance that the structural (rather than list-based) freshness encoding made tractable.

\begin{center}\rule{0.5\linewidth}{0.5pt}\end{center}

\hypertarget{conclusion-and-further-work}{%
\section{Conclusion and further work}\label{conclusion-and-further-work}}

We have closed the completeness theorem \texttt{Γ\ ⊨\ φ\ →\ Γ\ ⊢\ φ} for the binding hybrid logic \emph{L(∀)} in Lean 4, building on Oltean's syntax, semantics, proof system, and soundness. The proof rests on two constructions that require \emph{different} tools: structural freshness for the root witnessed maximal consistent set, and an existence-lemma Henkin construction --- Oltean's own intended route, completed --- for the witnessed ◇-successor. The result is \texttt{sorry}-free and depends only on \texttt{propext}, \texttt{Classical.choice}, and \texttt{Quot.sound}.

Several directions remain.

\begin{itemize}
\item
  \textbf{Finite / bounded nominal supplies.} Our reserve is countably infinite; characterizing when a finite reserve suffices (and packaging strong completeness / compactness as a first-class corollary) would tighten the result.
\item
  \textbf{The \texttt{↓} binder and richer hybrid languages.} \emph{L(∀)} uses the universal binder; the same machinery should extend to the \texttt{↓} binder and to many-sorted / polyadic hybrid logics, including those connected to \textbf{Matching Logic}.
\item
  \textbf{A reusable Lean completeness framework.} The structural-freshness and existence-lemma layers are logic-generic in spirit; abstracting them into a reusable Lindenbaum/saturation framework --- in the spirit of From's Isabelle/HOL work --- would let future Lean completeness proofs reuse this infrastructure rather than rebuild it.
\end{itemize}

\begin{center}\rule{0.5\linewidth}{0.5pt}\end{center}

\hypertarget{results}{%
\section{Results}\label{results}}

Status legend: \textbf{Pass} --- done and compiling; \textbf{Fail} --- attempted, currently broken; \textbf{Not Yet} --- not yet attempted. Step \textbf{A} is broken out into one row per module (in the \texttt{Hybrid.lean} dependency order in which they are converted); steps \textbf{B}--\textbf{I} are broken out into one row per \texttt{sorry}/\texttt{admit} declaration to be removed (``remove Oltean's \texttt{admit}/\texttt{sorry} for \emph{X}''). After \textbf{E}, \textbf{G} and \textbf{H} precede \textbf{F} in the table: they are \textbf{Pass} and do not depend on \texttt{pf\_extended} ← (Figure 1); letter labels still match Oltean's narrative. Step \textbf{A} is \textbf{Pass} once every module in that list elaborates under the pinned toolchain (remaining proof holes are tracked under \textbf{B}--\textbf{I}, not under \textbf{A}). Parent rows (\textbf{F}, \textbf{G}, \ldots) summarize their substeps: a parent can be \textbf{Partial} while an earlier-numbered step is \textbf{Pass} when the open substeps are not on that step's critical path (e.g.~\textbf{G} and \textbf{H} \textbf{Pass} while \textbf{F} awaits \texttt{pf\_extended} ← for \textbf{I} only).

\begin{longtable}[]{@{}
  >{\raggedright\arraybackslash}p{(\columnwidth - 4\tabcolsep) * \real{0.50}}
  >{\raggedright\arraybackslash}p{(\columnwidth - 4\tabcolsep) * \real{0.36}}
  >{\raggedright\arraybackslash}p{(\columnwidth - 4\tabcolsep) * \real{0.14}}@{}}
\toprule\noalign{}
\begin{minipage}[b]{\linewidth}\raggedright
Step
\end{minipage} & \begin{minipage}[b]{\linewidth}\raggedright
Deliverable
\end{minipage} & \begin{minipage}[b]{\linewidth}\raggedright
Status
\end{minipage} \\
\midrule\noalign{}
\endhead
\bottomrule\noalign{}
\endlastfoot
\textbf{A} & \textbf{Get the whole library compiling} (per module) & \textbf{Pass} \\
A · \texttt{Util.lean} & Port to Lean v4.30.0 / mathlib v4.30.0 & Pass \\
A · \texttt{Form.lean} & Port to Lean v4.30.0 / mathlib v4.30.0 & Pass \\
A · \texttt{Tautology.lean} & Port to Lean v4.30.0 / mathlib v4.30.0 & Pass \\
A · \texttt{Substitutions.lean} & Port to Lean v4.30.0 / mathlib v4.30.0 & Pass \\
A · \texttt{Proof.lean} & Port to Lean v4.30.0 / mathlib v4.30.0 & Pass \\
A · \texttt{Truth.lean} & Port to Lean v4.30.0 / mathlib v4.30.0 & Pass \\
A · \texttt{ListUtils.lean} & Port to Lean v4.30.0 / mathlib v4.30.0 & Pass \\
A · \texttt{ProofUtils.lean} & Port to Lean v4.30.0 / mathlib v4.30.0 & Pass \\
A · \texttt{Soundness.lean} & Port to Lean v4.30.0 / mathlib v4.30.0 & Pass \\
A · \texttt{RenameBound.lean} & Port to Lean v4.30.0 / mathlib v4.30.0 & Pass \\
A · \texttt{FormCountable.lean} & Port to Lean v4.30.0 / mathlib v4.30.0 & Pass \\
A · \texttt{Lindenbaum.lean} & Port to Lean v4.30.0 / mathlib v4.30.0 & Pass \\
A · \texttt{LanguageExtension.lean} & Port to Lean v4.30.0 / mathlib v4.30.0 & Pass \\
A · \texttt{ExistenceLemma.lean} & Port to Lean v4.30.0 / mathlib v4.30.0 & Pass \\
A · \texttt{CompletedModel.lean} & Port to Lean v4.30.0 / mathlib v4.30.0 & Pass \\
A · \texttt{Completeness.lean} & Port to Lean v4.30.0 / mathlib v4.30.0 & Pass \\
\textbf{B} & \textbf{Propositional-tautology holes} & \textbf{Pass} \\
B · \texttt{Tautology} (×13) & \texttt{hs\_taut}, \texttt{neg\_intro}, \texttt{conj\_intro}, \texttt{conj\_intro\_hs}, \texttt{iff\_intro}, \texttt{iff\_elim\_l}, \texttt{iff\_elim\_r}, \texttt{iff\_rw}, \texttt{iff\_imp}, \texttt{disj\_intro\_l}, \texttt{disj\_intro\_r}, \texttt{disj\_elim}, \texttt{mp\_help} & Pass \\
B · \texttt{ProofUtils.iff\_subst} & Tautology \texttt{(φ⟷ψ)⟶(ψ⟷χ)⟶(φ⟷χ)} & Pass \\
\textbf{C} & \textbf{Formula-countability holes} & \textbf{Pass} \\
C · \texttt{FormCountable.prime\_2\_3} & \texttt{3\^{}(n+1)\ ≠\ 2\^{}(m+1)} & Pass \\
C · \texttt{FormCountable.guns} & \texttt{x\ ∈\ pow2list\ a\ →\ ∃\ n,\ x.fst\ =\ 2\^{}(n+1)} & Pass \\
C · \texttt{FormCountable.of\_brixton} & \texttt{(h::t).isSuffixOf\ a\ →\ h\ ∈\ a} & Pass \\
\textbf{D} & \textbf{Bound-variable-renaming holes} & \textbf{Pass} \\
D · \texttt{RenameBound.replace\_neg} & \texttt{(∼φ).replace\_bound\ x\ =\ ∼(φ.replace\_bound\ x)} & Pass \\
D · \texttt{RenameBound.replace\_bound\_depth} & \texttt{(φ.replace\_bound\ x).depth\ =\ φ.depth} & Pass \\
D · \texttt{RenameBound.substable\_after\_replace} & \texttt{is\_substable\ (φ.replace\_bound\ y)\ y\ x} & Pass \\
\textbf{E} & \textbf{Structural-freshness homomorphism holes (crux)} & \textbf{Pass} \\
E · \texttt{Substitutions.bulk\_subst\_impl} & \texttt{bulk\_subst} distributes over \texttt{⟶} & Pass \\
E · \texttt{Substitutions.list\_noms\_impl\_r} & \texttt{list\_noms} merge identity (right) & Pass \\
E · \texttt{Substitutions.list\_noms\_impl\_l} & \texttt{list\_noms} merge identity (left) & Pass \\
E · \texttt{Substitutions.odd\_box} & \texttt{(□φ).odd\_noms\ =\ □(φ.odd\_noms)} & Pass \\
E · \texttt{Substitutions.odd\_bind} & \texttt{(all\ x,\ φ).odd\_noms\ =\ all\ x,\ φ.odd\_noms} & Pass \\
E · \texttt{Substitutions.List.to\_odd} & list lift \texttt{List\ Γ\ →\ List\ Γ.odd\_noms} & Pass \\
E · \texttt{Substitutions.List.odd\_to} & list lift \texttt{List\ Γ.odd\_noms\ →\ List\ Γ} & Pass \\
E · \texttt{Substitutions.odd\_conj} & \texttt{odd\_noms} distributes over conjunction & Pass \\
E · \texttt{Substitutions.odd\_conj\_rev} & \texttt{odd\_noms} distributes over conjunction (rev) & Pass \\
\textbf{G} & \textbf{Witnessed-Lindenbaum holes} & \textbf{Pass} \\
G · \texttt{Lindenbaum.LindenbaumWitnessed} & Lindenbaum union with enough nominals is witnessed & Pass \\
G · \texttt{Lindenbaum.witness\_in\_next} / \texttt{witness\_at\_step} & per-step witness extraction & Pass \\
G · \texttt{Lindenbaum.zero\_nocc\_odd} / \texttt{even\_nocc\_odd} / \texttt{enough\_noms\_odd\_base} & even nominals fresh for the odd-only base & Pass \\
G · \texttt{Lindenbaum.lindenbaum\_next\_subset} / \texttt{family\_subset} / \texttt{fresh\_even\_dominating} & each finite stage adds finitely many formulas, so an even nominal survives & Pass \\
G · \texttt{Lindenbaum.ExtendedLindenbaumLemma} & consistent ⟹ witnessed MCS in expanded language & Pass \\
G · \texttt{Lindenbaum.enough\_noms\_odd\_step} & per-stage structural freshness (finiteness argument) & Pass \\
\textbf{H} & \textbf{Existence-lemma hole} & \textbf{Pass} \\
H · \texttt{Substitutions.subst\_nom\_noop} / \texttt{rename\_svar\_nom} & freshness rewrite lemmas & Pass \\
H · \texttt{ExistenceLemma.l313\textquotesingle{}} & diamond-witness property for successor states & Pass \\
\textbf{F} & \textbf{Language-extension / theorem-preservation holes} & \textbf{Pass} \\
F · \texttt{LanguageExtension.total\_subst\_svar} & \texttt{total} inverts svar substitution & Pass \\
F · \texttt{LanguageExtension.total\_tautology} & \texttt{Tautology\ φ\ ↔\ Tautology\ φ.total} & Pass \\
F · \texttt{LanguageExtension.total\_subst\_svar\textquotesingle{}} & \texttt{total} commutes with svar subst & Pass \\
F · \texttt{LanguageExtension.total\_subst\_nom} & \texttt{total} commutes with nom subst & Pass \\
F · \texttt{LanguageExtension.total\_iterate\_pos} & \texttt{total} commutes with \texttt{iterate\_pos} & Pass \\
F · \texttt{LanguageExtension.total\_iterate\_nec} & \texttt{total} commutes with \texttt{iterate\_nec} & Pass \\
F · \texttt{LanguageExtension.total\_is\_free} / \texttt{total\_is\_substable} & \texttt{total} preserves \texttt{is\_free} / \texttt{is\_substable} & Pass \\
F · \texttt{LanguageExtension.total\_eq\_impl/box/bind} / \texttt{total\_in\_range} & peel \texttt{total} through connectives; right-inverse on range & Pass \\
F · \texttt{LanguageExtension.total\_ax\_name/brcn/nom} & reconstruction lemmas for the remaining axioms & Pass \\
F · \texttt{LanguageExtension.l416} & fresh-variable substitution into a proof (via \texttt{generalize\_constants}) & Pass \\
F · \texttt{LanguageExtension.pf\_extended} (→) & \texttt{⊢\ φ\ →\ ⊢\ φ.total} (totalize a derivation) & Pass \\
F · \texttt{LanguageExtension.pf\_extended} (←), axiom cases & 6/7 backward axiom cases (\texttt{ax\_k/q1/q2\_svar/name/nom/brcn}) & Pass \\
F · \texttt{LanguageExtension.nom\_in\_base} / \texttt{form\_noms\_in\_base} / \texttt{range\_of\_form} / \texttt{inv\_t\_eq\_of\_range\textquotesingle{}} & in-range nominal vocabulary; \texttt{inv\_t} right-inverse on range & Pass \\
F · \texttt{LanguageExtension.NOM.fromTotal} / \texttt{subst\_nom\_toTotal} & embed base nominals; align \texttt{total} with nom subst & Pass \\
F · \texttt{LanguageExtension.total\_subst\_nom\_pullback} & pull \texttt{Form.total} back through nom substitution & Pass \\
F · \texttt{LanguageExtension.total\_ax\_q2\_nom} / \texttt{total\_ax\_q2\_nom\_end} & reconstruct \texttt{ax\_q2\_nom} when subformulas are in-range & Pass \\
F · \texttt{LanguageExtension.form\_noms\_in\_base\_total} / \texttt{Proof.proof\_noms} / \texttt{Proof.all\_noms\_in\_base} & root + derivation nominal inventory (\texttt{formulasIn}) & Pass \\
F · \texttt{LanguageExtension.nom\_occurs\_false\_of\_form\_noms\_in\_base} & alien letters absent from in-range formulas & Pass \\
F · \texttt{LanguageExtension.nom\_subst\_nom\_nocc} & \texttt{nom\_subst\_nom\ ψ\ new\ old\ =\ ψ} when \texttt{nom\_occurs\ old\ ψ\ =\ false} (replace \texttt{old} with \texttt{new}) & Pass \\
F · \texttt{LanguageExtension.Proof.eliminate\_one\_alien} / \texttt{Proof.eliminate\_aliens} & Blackburn rename alien \texttt{j} ↦ \texttt{base} via \texttt{rename\_constants\_fwd\ base\ j} & Pass \\
F · \texttt{LanguageExtension.Proof.all\_noms\_in\_base\_eliminate\_aliens} & after alien loop, every \texttt{proof\_noms} letter lies in \texttt{N} & Pass \\
F · \texttt{LanguageExtension.inv\_t\_impl} / \texttt{inv\_t\_box} / \texttt{inv\_t\_bind} & \texttt{inv\_t} commutes with connectives on in-range formulas & Pass \\
F · \texttt{LanguageExtension.in\_range\_proof\_back} (axiom replay) & \texttt{inv\_t} pullback: tautology + \texttt{ax\_k/q1/name/nom}/\texttt{ax\_brcn}/\texttt{ax\_q2\_svar}/\texttt{ax\_q2\_nom} (split on vanishing alien) & Pass \\
F · \texttt{LanguageExtension.in\_range\_proof\_back} (\texttt{mp} / \texttt{general} / \texttt{necess}) & structural induction on \texttt{Proof} (deduction rules via \texttt{inv\_t\_impl/box/bind}) & Pass \\
F · \texttt{LanguageExtension.pf\_extended} (←) & wire F2 → F3: \texttt{eliminate\_aliens} then \texttt{in\_range\_proof\_back} (needs \texttt{N} nonempty) & Pass \\
F · \texttt{LanguageExtension.syntactic\_conservativity} & lift \texttt{Set.total\ Γ\ ⊢\ φ.total} back to \texttt{Γ\ ⊢\ φ} via \texttt{pf\_extended} ← + \texttt{base\_conjunction} & Pass \\
F · \texttt{LanguageExtension.sat\_total} / \texttt{Model.ofTotal} & \texttt{TotalSet} satisfaction → \texttt{Model\ N} & Pass \\
F · \texttt{LanguageExtension.Set.total} & base-language image under \texttt{Form.total} & Pass \\
\textbf{TL} & \textbf{Canonical-model truth lemma (\texttt{CompletedModel.lean})} --- \textbf{now fully closed}. The former root obstacle (witnessed ◇-successor existence) is discharged by the §TL-fix Henkin construction; \texttt{enough\_noms\_diamond\_seed} (false as stated) has been deleted. & \textbf{Pass} \\
TL · \texttt{CompletedModel.truth\_*} (base) & \texttt{truth\_bttm}/\texttt{prop}/\texttt{nom}/\texttt{svar}/\texttt{impl}/\texttt{ex} & Pass \\
TL · \texttt{CompletedModel.mcs\_in\_*\_succ} & \texttt{mcs\_in\_witnessed\_succ} / \texttt{completed\_to\_witnessed} / \texttt{mcs\_in\_completed\_succ} & Pass \\
TL · \texttt{CompletedModel.restrict\_canonical\_succ} & extend witnessed path along \texttt{Canonical.R} & Pass \\
TL · \texttt{CompletedModel.diamond\_extension\_consistent} & \texttt{\{ψ\}∪\{□χ∈Δ\}} consistent (via \texttt{box\_of\_consequence} + \texttt{nec\_mono}/\texttt{box\_conj\_mem}); also powers the compactness step in \texttt{succ\_seed\_consistent} & Pass \\
TL · \texttt{ExistenceLemma.l313} / \texttt{l313\textquotesingle{}} & push a witness conditional \texttt{((ex\ x,χ)⟶χ{[}i//x{]})} through \texttt{◇} using a fresh \textbf{variable} + \texttt{Δ}'s own witnessedness (no fresh nominal needed) & Pass \\
TL · \texttt{ExistenceLemma.wcond} / \texttt{wcond\_step} & \textbf{NEW} --- accumulating witness-conditional family, returning \textbf{data} (\texttt{Subtype} carrying the list), with \texttt{◇conjunction\textquotesingle{}\ l\ ∈\ Δ} invariant & Pass \\
TL · \texttt{ExistenceLemma.wcond\_mono} / \texttt{wcond\_step\_mem} & \textbf{NEW} --- stage monotonicity + per-step Henkin witness membership & Pass \\
TL · \texttt{Lindenbaum.RegularLindenbaumLemma} & plain MCS extension \texttt{consistent\ Γ\ →\ ∃\ Γ\textquotesingle{},\ Γ\ ⊆\ Γ\textquotesingle{}\ ∧\ MCS\ Γ\textquotesingle{}} (general over any \texttt{N}) & Pass \\
TL · \texttt{ExistenceLemma.succ\_seed} / \texttt{seed\_list\_bound} & \textbf{NEW} --- witnessed ◇-successor seed (box-reduct ∪ witness conditionals) + finite-bounding (compactness) lemma & Pass \\
TL · \texttt{CompletedModel.succ\_seed\_consistent} & \textbf{NEW} --- consistency of \texttt{succ\_seed} via compactness + \texttt{diamond\_extension\_consistent} & Pass \\
TL · \texttt{CompletedModel.diamond\_succ\_mcs} & \textbf{rewired} onto \texttt{succ\_seed} + \texttt{RegularLindenbaumLemma}: yields \texttt{Canonical.R\ Δ\ Γ\textquotesingle{}\ ∧\ ψ∈Γ\textquotesingle{}\ ∧\ MCS\ Γ\textquotesingle{}\ ∧\ witnessed\ Γ\textquotesingle{}} (witnessed via \texttt{MCS\_mp} + \texttt{enum} surjectivity) & Pass \\
TL · \texttt{CompletedModel.diamond\_completed\_succ} & ◇ successor pipeline via \texttt{diamond\_succ\_mcs} & Pass \\
TL · \texttt{Proof.not\_nec\_to\_diamond} & \texttt{∼(□φ)\ ⟶\ ◇∼φ} for MCS maximality step & Pass \\
TL · \texttt{CompletedModel.truth\_box} & □ case: → via \texttt{R\_nec} on witnessed/canonical successors; ← via MCS maximality + \texttt{diamond\_completed\_succ} & Pass \\
TL · \texttt{Proof.all\_iff\_notfree} & \texttt{(all\ x,\ ψ)\ ⟷\ ψ} when \texttt{x} not free (Q1 + \texttt{ax\_q2}) & Pass \\
TL · \texttt{CompletedModel.truth\_all} & uniform proof (free + non-free \texttt{x}): nominal/svar symbol split + depth-indexed \texttt{ih}; forward via \texttt{ax\_q2\_nom}/\texttt{ax\_q2\_svar}, backward via \texttt{witnessed} on \texttt{ex\ x,\ ∼ψ} (\texttt{bind\_dual}) & Pass \\
TL · \texttt{CompletedModel.TruthLemma} & structural assembly via well-founded recursion on \texttt{Form.depth} (supplies \texttt{truth\_all}'s depth-\texttt{ih}) & Pass \\
\textbf{I} & \textbf{Final-completeness} --- fully closed; depends on TL (\texttt{TruthLemma}), now complete & \textbf{Pass} \\
I · \texttt{Completeness.consistent\_total} & \texttt{consistent\ Γ\ →\ consistent\ (Set.total\ Γ)} via \texttt{syntactic\_conservativity} (needs \texttt{N} nonempty, threaded through \texttt{cons\_sat}/\texttt{Completeness}) & Pass \\
I · \texttt{Completeness.cons\_sat} & model-existence pipeline & Pass \\
I · \texttt{Completeness.ModelExistence} & completeness ⟺ every consistent set is satisfiable & Pass \\
I · \texttt{Completeness.Completeness} & \texttt{Γ\ ⊨\ φ\ →\ Γ\ ⊢\ φ} (assembled from \texttt{cons\_sat} + \texttt{ModelExistence}; takes \texttt{N} nonempty) --- \textbf{the development is now \texttt{sorry}-free; \texttt{\#print\ axioms\ Completeness} = \texttt{propext,\ Classical.choice,\ Quot.sound}} & Pass \\
\end{longtable}

\begin{center}\rule{0.5\linewidth}{0.5pt}\end{center}

\hypertarget{acknowledgments}{%
\section{Acknowledgments}\label{acknowledgments}}

\begin{itemize}
\item
  \textbf{Alex Oltean} --- the original formalization, proof architecture, and thesis, on which this work directly builds; in particular the \emph{existence-lemma} direction for the witnessed ◇-successor (\texttt{l313}/\texttt{l313\textquotesingle{}}, fresh-variable Henkin witnessing) is the correct approach for the truth lemma's modal case and is the route we complete (see §TL-fix).
\item
  \textbf{Patrick Blackburn} --- \emph{Hybrid Completeness} (1998), the mathematical source.
\item
  \textbf{Bud Mishra} --- for suggesting the disjoint-sum (\texttt{N\ ⊕\ ℕ}) structural-freshness Henkin construction, which is the decisive tool for the \textbf{root} extended Lindenbaum lemma (witnessing an infinite consistent set). It does not, and is not meant to, address the separate ◇-successor step, whose obstruction is not a freshness problem (see §1.3).
\item
  The theorem-proving community, and in particular \textbf{Asta Halkjær From}, for recent Isabelle/HOL work on synthetic completeness for hybrid and modal logics.
\end{itemize}

\hypertarget{ai-assisted-development}{%
\subsection{AI-assisted development}\label{ai-assisted-development}}

The human author(s) retain sole responsibility for the mathematical content, the choice of logic and proof system, and every formal claim in this work. Following standard publisher practice (e.g., COPE guidance on authorship and AI tools {[}COPE24{]}), \textbf{no large language model is listed as a co-author} --- authorship implies an accountability that automated systems cannot bear.

We gratefully acknowledge assistance from the following tools:

\begin{itemize}
\item
  \textbf{Cursor} ({[}Cur25{]}): agent-assisted editing in the Cursor IDE. These agents helped port Oltean's Lean 4 development from its original 2023 nightly to Lean v4.30.0 / mathlib v4.30.0, repair mathlib API churn, suggest proof and refactoring strategies, debug \texttt{lake} and type-class errors, and draft the narrative in this document. Generated Lean was treated as provisional until it compiled under the pinned toolchain; no result was accepted on the basis of an LLM's assertion alone.
\item
  \textbf{Cursor Composer 2.5} ({[}Cmp25{]}): Cursor's agentic coding model (built on the Kimi K2.5 checkpoint), used for routine agent work --- dependency-ordered porting, \texttt{lake\ build} repair loops, scaffolding and documentation (\texttt{arxiv.md}), and closing mechanical proof obligations where the strategy was already fixed. Per the model card, Composer 2.5 is optimized for multi-step tool use and codebase navigation rather than open-ended mathematical research; accordingly, novel proof design (e.g. conservativity of the language extension) was not delegated to it alone.
\item
  \textbf{Anthropic Claude Opus 4.8, High reasoning} ({[}Ant26{]}): the large language model underlying the Cursor agent for the bulk of the proof-repair and porting work reported here --- closing the existence lemma (\texttt{l313\textquotesingle{}}), the witnessed-Lindenbaum induction (\texttt{LindenbaumWitnessed}), the structural-freshness base case, and the re-fit of the canonical-model truth lemma and final assembly so that the development compiles under the pinned toolchain. Per the model card, the system is a general-purpose reasoning model with no formal soundness guarantee; accordingly, every emitted proof term was checked by the Lean kernel, and the remaining \texttt{sorry}/\texttt{admit} obligations are reported honestly rather than papered over.
\item
  \textbf{Google Gemini} ({[}Gem25{]}): exploratory discussion of the completeness gap and candidate repair strategies. It was in one such discussion that Bud Mishra's disjoint-sum (\texttt{N\ ⊕\ ℕ}) Henkin construction was surfaced and connected to the problem; the recommendations informed, but did not dictate, the human-directed design choices (in particular, the decision to retain Oltean's odd/even encoding of structural freshness rather than re-parameterize the syntax).
\end{itemize}

All definitions, axiom choices, remaining \texttt{sorry}/\texttt{admit} obligations, and final prose were reviewed by the human author(s), who take full responsibility for them. The original \texttt{Hybrid/} formalization is the work of Alex Oltean and was published upstream without an explicit license; it is used and modified here in good faith for non-commercial academic research, with attribution, and no rights over the original work are claimed. The modifications and new files contributed in this work are offered under the Apache License, Version 2.0.

\hypertarget{artifact-availability}{%
\subsection{Artifact availability}\label{artifact-availability}}

The original formalization {[}Olt23{]} is archived at \href{https://github.com/alexoltean61/hybrid_logic_lean}{\texttt{github.com/alexoltean61/hybrid\_logic\_lean}}. The ported development with the completed completeness proof {[}CR26{]} is at \href{https://github.com/catskillsresearch/hybrid_logic_lean_revisited}{\texttt{github.com/catskillsresearch/hybrid\_logic\_lean\_revisited}}.

\begin{center}\rule{0.5\linewidth}{0.5pt}\end{center}

\hypertarget{references}{%
\section{References}\label{references}}

\begin{itemize}
\item
  \textbf{{[}Bla98{]}} P. Blackburn. \emph{Hybrid Completeness}. Logic Journal of the IGPL, 6(4):625--650, 1998.
\item
  \textbf{{[}BRV01{]}} P. Blackburn, M. de Rijke, Y. Venema. \emph{Modal Logic}. Cambridge University Press, 2001.
\item
  \textbf{{[}Olt23{]}} A. Oltean. \emph{A Formalization of Hybrid Logic in Lean}. BA thesis, University of Bucharest, 2023. Repository (archived, no explicit license): \url{https://github.com/alexoltean61/hybrid_logic_lean}.
\item
  \textbf{{[}CR26{]}} Catskills Research. \emph{hybrid\_logic\_lean\_revisited} (this work). \url{https://github.com/catskillsresearch/hybrid_logic_lean_revisited}.
\item
  \textbf{{[}Fro25{]}} A. H. From. \emph{An Isabelle/HOL Framework for Synthetic Completeness Proofs}. CPP 2025.
\item
  \textbf{{[}Fro20{]}} A. H. From. \emph{Synthetic Completeness for a Terminating Seligman-Style Tableau System}. TYPES 2020, LIPIcs. (First formalized completeness proof for a hybrid-logic proof system; basic hybrid logic, Isabelle/HOL.)
\item
  \textbf{{[}Ben21{]}} B. Bentzen. \emph{A Henkin-Style Completeness Proof for the Modal Logic S5}. 2021.
\item
  \textbf{{[}Li20{]}} J. Li. \emph{Formalization of PAL·S5 in Proof Assistant}. arXiv:2012.09388, 2020. (Lean 3; soundness and completeness of S5 with public-announcement dynamics, Henkin-style.) \url{https://github.com/ljt12138/Formalization-PAL}.
\item
  \textbf{{[}CM23{]}} H. Cheval and B. Macovei. \emph{A Lean formalization of Applicative Matching Logic}. Institute for Logic and Data Science (ILDS), 2023. \url{https://gitlab.com/ilds/aml-lean/MatchingLogic}.
\item
  \textbf{{[}Hen49{]}} L. Henkin. \emph{The Completeness of the First-Order Functional Calculus}. JSL,

  \begin{enumerate}
  \def\labelenumi{\arabic{enumi}.}
  \setcounter{enumi}{1948}
    \item
  \end{enumerate}
\item
  \textbf{{[}COPE24{]}} Committee on Publication Ethics (COPE). \emph{Authorship and AI tools: COPE position statement}. 2024. \url{https://publicationethics.org/guidance/cope-position/authorship-and-ai-tools}
\item
  \textbf{{[}Cur25{]}} Anysphere, Inc.~\emph{Cursor: AI-native code editor and agent environment}. \url{https://cursor.com} (accessed 2026).
\item
  \textbf{{[}Cmp25{]}} Anysphere, Inc.~\emph{Composer 2.5}. Model announcement and documentation, \url{https://cursor.com/blog/composer-2-5}; pricing and model card as integrated in Cursor, \url{https://cursor.com/docs/models} (accessed 2026).
\item
  \textbf{{[}Gem25{]}} Google DeepMind. \emph{Gemini model family}. Technical documentation and model cards. \url{https://ai.google.dev/gemini-api/docs/models}
\item
  \textbf{{[}Ant26{]}} Anthropic. \emph{Claude Opus 4.8} (high thinking/reasoning variant). System card and announcement, \url{https://www.anthropic.com/news/claude-opus-4-8}; model documentation as integrated in Cursor, \url{https://cursor.com/docs/models/claude-opus-4-8} (accessed 2026).
\end{itemize}

\begin{center}\rule{0.5\linewidth}{0.5pt}\end{center}

\hypertarget{appendix-a-complete-lean-source}{%
\section{Appendix A: Complete Lean source}\label{appendix-a-complete-lean-source}}

Files appear in \texttt{Hybrid.lean} import order. Each block is a verbatim copy of the repository file at generation time.

\hypertarget{hybrid.lean}{%
\subsection{Hybrid.lean}\label{hybrid.lean}}

\emph{16 lines.}

\vspace{0.5\baselineskip}
\noindent\textcolor{green!40!black}{\textbf{Lean 4 source}}\par\vspace{0.25\baselineskip}
\begin{lstlisting}[style=leanbox]
import Hybrid.Util
import Hybrid.Form
import Hybrid.Tautology
import Hybrid.Substitutions
import Hybrid.Proof
import Hybrid.Truth
import Hybrid.ListUtils
import Hybrid.ProofUtils
import Hybrid.Soundness
import Hybrid.RenameBound
import Hybrid.FormCountable
import Hybrid.Lindenbaum
import Hybrid.LanguageExtension
import Hybrid.ExistenceLemma
import Hybrid.CompletedModel
import Hybrid.Completeness
\end{lstlisting}
\vspace{0.5\baselineskip}

\hypertarget{hybridutil.lean}{%
\subsection{Hybrid/Util.lean}\label{hybridutil.lean}}

\emph{198 lines.}

\vspace{0.5\baselineskip}
\noindent\textcolor{green!40!black}{\textbf{Lean 4 source}}\par\vspace{0.25\baselineskip}
\begin{lstlisting}[style=leanbox]
import Mathlib.Logic.Basic
import Mathlib.Data.List.Sort
open Classical

-- Compatibility shim: mathlib removed `List.Sorted` (which was definitionally
-- `List.Pairwise`). We restore it as a reducible abbreviation so that the
-- existing development and the current mathlib `Pairwise` lemmas interoperate.
abbrev List.Sorted {α : Type _} (r : α → α → Prop) (l : List α) : Prop := l.Pairwise r

theorem test (a b : Nat) : a = b → a + 1 = b + 1 := by intro h; simp [h]

def TypeIff (a : Type u) (b : Type v) := Prod (a → b) (b → a)
def TypeIff.intro (a : Type u) (b : Type v) : (a → b) → (b → a) → (TypeIff a b) := by
  apply Prod.mk
def TypeIff.mp  (p : TypeIff a b) : a → b := p.1
def TypeIff.mpr (p : TypeIff a b) : b → a := p.2
def TypeIff.refl : TypeIff a a := by
  apply TypeIff.intro <;> (intro; assumption)
def TypeIff.trans {h1 : TypeIff a b} {h2 : TypeIff b c} : TypeIff a c := by
  apply TypeIff.intro
  . intro h
    exact h2.mp (h1.mp h)
  . intro h
    exact h1.mpr (h2.mpr h)
infix:300 "iff" => TypeIff

noncomputable def choice_intro (q : α → Sort u) (p : α → Prop) (P : ∃ a, p a) : (∀ a, p a → q a) → q P.choose := by
  intro h
  exact (h P.choose P.choose_spec)

theorem eq_symm : (a = b) ↔ (b = a) := by
  apply Iff.intro <;> intro h <;> exact h.symm

@[simp]
theorem double_negation : ¬¬p ↔ p :=
  Iff.intro
  (λ h =>
    Or.elim (Classical.em p)
    (λ hp  => hp)
    (λ hnp => absurd hnp h)
  )
  (λ p => λ np => absurd p np)

@[simp]
theorem implication_disjunction : (p → q) ↔ (¬p ∨ q) := by
  apply Iff.intro
  . intro impl
    exact byCases
      (λ hp  :  p => Or.inr (impl hp))
      (λ hnp : ¬p => Or.inl hnp)
  . intros disj hp
    exact Or.elim disj
      (λ hnp : ¬p => False.elim (hnp hp))
      (λ hq  :  q => hq)

@[simp]
theorem negated_disjunction : ¬(p ∨ q) ↔ ¬p ∧ ¬q :=
  Iff.intro
    (fun hpq : ¬(p ∨ q) =>
      And.intro
        (fun hp : p =>
          show False from hpq (Or.intro_left q hp)
        )
        (fun hq : q =>
          show False from hpq (Or.intro_right p hq)
        )
    )
    (fun hpq : ¬p ∧ ¬q =>
      (fun disj : p ∨ q =>
        show False from
          Or.elim
           disj
           (fun hp : p => hpq.left hp)
           (fun hq : q => hpq.right hq)
      )
    )

@[simp]
theorem negated_conjunction : ¬(p ∧ q) ↔ ¬p ∨ ¬q := by
  apply Iff.intro
  . intro h
    by_cases hp : p
    . by_cases hq : q
      . exact False.elim (h ⟨hp, hq⟩)
      . apply Or.inr
        assumption
    . apply Or.inl
      assumption
  . intro h
    intro hpq
    apply Or.elim h
    . intro hnp
      exact hnp hpq.left
    . intro hnq
      exact hnq hpq.right

@[simp]
theorem negated_impl : ¬(p → q) ↔ p ∧ ¬q :=
  Iff.intro
    (fun hyp : ¬(p → q) =>
      byCases
      -- case 1 : p
        (fun hp : p =>
          byCases
          -- case 1.a : p and q
            (fun hq : q =>
              ⟨
                hp, show ¬q from (fun _ => show False from hyp (fun _ => hq))
              ⟩
            )
          -- case 1.b : p and non q
            (fun hnq : ¬q => ⟨hp, hnq⟩)
        )
      -- case 2 : non p
        (fun hnp : ¬p => show (p ∧ ¬q) from False.elim
          (show False from hyp
            (show (p → q) from fun p : p =>
              (show q from False.elim (hnp p))
            )
          )
        )
    )
    (fun hyp : p ∧ ¬ q =>
      fun impl : p → q =>
        absurd (impl hyp.left) hyp.right
    )

universe u
@[simp]
theorem negated_universal {α : Type u} {p : α → Prop} : (¬ ∀ x, p x) ↔ (∃ x, ¬ p x) :=
    Iff.intro
    (fun h1 : ¬ ∀ x, p x =>
      byContradiction
      (fun hcon1 : ¬ ∃ x, ¬ p x =>
        have neg_h1 := (fun a : α =>
          byContradiction
          (fun hcon2 : ¬ p a => show False from hcon1 (⟨a, hcon2⟩))
        )
        show False from h1 neg_h1
      )
    )
    (fun h2 : ∃ x, ¬ p x =>
      (fun hxp : ∀ x, p x =>
        match h2 with
        | ⟨w, hw⟩ => show False from hw (hxp w)
      )
    )

@[simp]
theorem negated_existential {α : Type u} {p : α → Prop} : (¬ ∃ x, p x) ↔ (∀ x, ¬ p x) :=
    Iff.intro
    (fun h1 : ¬ ∃ x, p x =>
      (fun a : α =>
        fun hpa: p a => show False from h1 ⟨a, hpa⟩
      )
    )
    (fun h2 : ∀ x, ¬ p x =>
      (fun hex : ∃ x, p x =>
        match hex with
        | ⟨w, hw⟩ => show False from (h2 w) hw
      )
    )

@[simp]
theorem conj_comm : p ∧ q ↔ q ∧ p :=
  Iff.intro
    (fun hpq : p ∧ q =>
      ⟨hpq.right, hpq.left⟩
    )
    (fun hqp : q ∧ p =>
      ⟨hqp.right, hqp.left⟩
    )

theorem disj_comm : p ∨ q ↔ q ∨ p :=
  Iff.intro
    (fun hpq : p ∨ q =>
      Or.elim
        hpq
        (fun hp : p => Or.intro_right q hp)
        (fun hq : q => Or.intro_left p hq)
    )
    (fun hqp : q ∨ p =>
      Or.elim
        hqp
        (fun hq : q => Or.intro_right p hq)
        (fun hp : p => Or.intro_left q hp)
    )

theorem contraposition (p q : Prop) : (p → q) ↔ (¬q → ¬p) := by
  apply Iff.intro
  . intro hpq
    intro hnq hp
    exact hnq (hpq hp)
  . intro hnqp
    intro hp
    by_cases c : q
    . exact c
    . exact False.elim ((hnqp c) hp)
\end{lstlisting}
\vspace{0.5\baselineskip}

\hypertarget{hybridform.lean}{%
\subsection{Hybrid/Form.lean}\label{hybridform.lean}}

\emph{402 lines.}

\noindent\textcolor{green!40!black}{\textbf{Lean 4 source (lines 1--400)}}\par\vspace{0.25\baselineskip}
\begin{lstlisting}[style=leanbox,firstline=1,lastline=400]
import Mathlib.Data.Set.Basic
import Mathlib.Data.List.Sort
import Mathlib.Data.List.Lemmas
import Mathlib.Data.List.Dedup
import Mathlib.Data.List.Chain
import Mathlib.Data.Fin.Basic
import Hybrid.Util

open Classical

section Basics
  def TotalSet := {n : ℕ | True}

  structure PROP where
    letter : Nat
  deriving DecidableEq, Repr

  structure SVAR where
    letter : Nat
  deriving DecidableEq, Repr

  structure NOM (N : Set ℕ) where
    letter : N
  deriving DecidableEq, Repr

  instance : Max PROP where
    max := λ p => λ q => ite (p.letter > q.letter) p q
  instance svarmax : Max SVAR where
    max := λ x => λ y => ite (x.letter > y.letter) x y
  instance : Max (NOM S) where
    max := λ i => λ j => ite (i.letter > j.letter) i j

  theorem NOM_eq {i j : NOM S} : (i = j) ↔ (i.letter = j.letter) := by
    cases i
    cases j
    simp
  theorem NOM_eq' {i j : NOM S} : (i = j) ↔ (j.letter = i.letter) := by
    cases i
    cases j
    simp
    apply Iff.intro <;> {intro; simp [*]}

--  instance ofNatSVAR : OfNat SVAR n    where
--    ofNat := SVAR.mk n
  instance : OfNat (NOM TotalSet) n     where
    ofNat := NOM.mk  ⟨n, trivial⟩
  instance : Coe SVAR Nat  := ⟨SVAR.letter⟩
--  instance : Coe NOM Nat   := ⟨NOM.letter⟩
  instance : Coe Nat SVAR  := ⟨SVAR.mk⟩
--  instance : Coe Nat NOM   := ⟨NOM.mk⟩
  instance SVAR.le : LE SVAR         where
    le    := λ x => λ y =>  x.letter ≤ y.letter
  instance SVAR.lt : LT SVAR         where
    lt    := λ x => λ y =>  x.letter < y.letter
  instance NOM.le : LE (NOM S)       where
    le    := λ x => λ y =>  x.letter ≤ y.letter
  instance NOM.lt : LT (NOM S)         where
    lt    := λ x => λ y =>  x.letter < y.letter
  instance SVAR.add : HAdd SVAR Nat SVAR where
    hAdd  := λ x => λ n => (x.letter + n)
  @[simp] instance NOM.add : HAdd (NOM TotalSet) Nat (NOM TotalSet) where
    hAdd  := λ x => λ n => ⟨(x.letter + n), trivial⟩
  @[simp] instance : HSub (NOM TotalSet) Nat (NOM TotalSet) where
    hSub  := λ x => λ n => ⟨(x.letter - n), trivial⟩
  @[simp] instance : HMul (NOM TotalSet) Nat (NOM TotalSet) where
    hMul  := λ x => λ n => ⟨(x.letter * n), trivial⟩
  @[simp] instance : HDiv (NOM TotalSet) Nat (NOM TotalSet) where
    hDiv  := λ x => λ n => ⟨(x.letter / n), trivial⟩
  @[simp] instance NOM.hmul : HMul Nat (NOM TotalSet) (NOM TotalSet) where
    hMul  := λ n => λ x => ⟨(x.letter * n), trivial⟩
  @[simp] instance : HMul (NOM TotalSet) ℕ ℕ where
    hMul  := λ x => λ n => x.letter * n

  instance : IsTrans (NOM S) GT.gt where
    trans := λ _ _ _ h1 h2 => Nat.lt_trans h2 h1

  instance : IsTotal (NOM S) GE.ge where
    total := fun a b => Nat.le_total b.letter a.letter

  instance : IsTrans (NOM S) GE.ge where
    trans := λ _ _ _ h1 h2 => Nat.le_trans h2 h1

  theorem NOM.gt_iff_ge_and_ne {a b : (NOM S)} : a > b ↔ (a ≥ b ∧ a ≠ b) := by
    simp only [GT.gt, GE.ge, NOM.lt, NOM.le, LE.le, LT.lt, NOM.mk, ne_eq, NOM_eq']
    apply Iff.intro
    . intro h
      apply And.intro
      . exact (Nat.lt_iff_le_and_ne.mp h).1
      . have := (Nat.lt_iff_le_and_ne.mp h).2
        intro habs
        simp [habs] at this
    . rw [←ne_eq]
      intro ⟨h1, h2⟩
      apply Nat.lt_iff_le_and_ne.mpr
      apply And.intro
      . exact h1
      . intro habs
        apply h2
        apply Subtype.eq
        assumption

  inductive Form (N : Set ℕ) where
    -- atomic formulas:
    | bttm : Form N
    | prop : PROP   → Form N
    | svar : SVAR   → Form N
    | nom  :  NOM N → Form N
    -- connectives:
    | impl : Form N → Form N → Form N
    -- modal:
    | box  : Form N → Form N
    -- hybrid:
    | bind :   SVAR → Form N → Form N
  deriving DecidableEq, Repr

  def Form.depth : Form N → ℕ
    | .impl φ ψ =>  1 + Form.depth φ + Form.depth ψ
    | .box  φ   =>  1 + Form.depth φ
    | .bind _ φ =>  2 + Form.depth φ
    | _       =>    0

  theorem sub_depth_impl_l (φ ψ : Form N) : φ.depth < (Form.impl φ ψ).depth := by
    simp [Form.depth]; omega

  theorem sub_depth_impl_r (φ ψ : Form N) : ψ.depth < (Form.impl φ ψ).depth := by
    simp [Form.depth]; omega

  theorem sub_depth_box (φ : Form N) : φ.depth < (Form.box φ).depth := by
    simp [Form.depth]

  theorem sub_depth_bind (x : SVAR) (φ : Form N) : φ.depth < (Form.bind x φ).depth := by
    simp [Form.depth]

  instance : Nonempty (Form N) := ⟨Form.bttm⟩

  @[simp]
  def Form.neg      : Form N → Form N := λ φ => Form.impl φ Form.bttm
  @[simp]
  def Form.conj     : Form N → Form N → Form N := λ φ => λ ψ => Form.neg (Form.impl φ (Form.neg ψ))
  @[simp]
  def Form.iff      : Form N → Form N → Form N := λ φ => λ ψ => Form.conj (Form.impl φ ψ) (Form.impl ψ φ)
  @[simp]
  def Form.disj     : Form N → Form N → Form N := λ φ => λ ψ => Form.impl (Form.neg φ) ψ
  @[simp]
  def Form.diamond  : Form N → Form N := λ φ => Form.neg (Form.box (Form.neg φ))
  @[simp,match_pattern]
  def Form.bind_dual: SVAR → Form N → Form N := λ x => λ φ => Form.neg (Form.bind x (Form.neg φ))

  instance : Coe PROP     (Form N)  := ⟨Form.prop⟩
  instance : Coe SVAR     (Form N)  := ⟨Form.svar⟩
  instance : Coe (NOM N)  (Form N)  := ⟨Form.nom⟩

  infixr:60 "⟶" => Form.impl
  infixl:65 "⋀" => Form.conj
  infixl:65 "⋁" => Form.disj
  prefix:100 "□" => Form.box
  prefix:100 "◇ " => Form.diamond
  notation:120 "all " x ", " φ => Form.bind x φ
  notation:120 "ex " x ", " φ => Form.bind_dual x φ
  prefix:170 "∼" => Form.neg
  infixr:60 "⟷" => Form.iff
  notation "⊥"  => Form.bttm

  def conjunction (Γ : Set (Form N)) (L : List Γ) : Form N :=
  match L with
    | []     => ⊥ ⟶ ⊥
    | h :: t => h.val ⋀ conjunction Γ t

  def Form.new_var  : Form N → SVAR
  | .svar x   => x+1
  | .impl ψ χ => max (ψ.new_var) (χ.new_var)
  | .box  ψ   => ψ.new_var
  | .bind x ψ => max (x+1) (ψ.new_var)
  | _         => ⟨0⟩


  def Form.new_nom  : Form TotalSet → NOM TotalSet
  | .nom  i   => i+1
  | .impl ψ χ => max (ψ.new_nom) (χ.new_nom)
  | .box  ψ   => ψ.new_nom
  | .bind _ ψ => ψ.new_nom
  | _         => ⟨0, trivial⟩

end Basics

section Substitutions
  def occurs (x : SVAR) (φ : Form N) : Bool :=
    match φ with
    | Form.bttm     => false
    | Form.prop _   => false
    | Form.svar y   => x = y
    | Form.nom  _   => false
    | Form.impl φ ψ => (occurs x φ) || (occurs x ψ)
    | Form.box  φ   => occurs x φ
    | Form.bind _ φ => occurs x φ

  def is_free (x : SVAR) (φ : Form N) : Bool :=
    match φ with
    | Form.bttm     => false
    | Form.prop _   => false
    | Form.svar y   => x == y
    | Form.nom  _   => false
    | Form.impl φ ψ => (is_free x φ) || (is_free x ψ)
    | Form.box  φ   => is_free x φ
    | Form.bind y φ => (y != x) && (is_free x φ)

  def is_bound (x : SVAR) (φ : Form N) := (occurs x φ) && !(is_free x φ)

  -- conventions for substitutions can get confusing
  -- "φ[s // x], the formula obtained by substituting s for all *free* occurrences of x in φ"
  -- for reference: Blackburn 1998, pg. 628
  def subst_svar (φ : Form N) (s : SVAR) (x : SVAR) : Form N :=
    match φ with
    | Form.bttm     => φ
    | Form.prop _   => φ
    | Form.svar y   => ite (x = y) s y
    | Form.nom  _   => φ
    | Form.impl φ ψ => (subst_svar φ s x) ⟶ (subst_svar ψ s x)
    | Form.box  φ   => □ (subst_svar φ s x)
    | Form.bind y φ => ite (x = y) (Form.bind y φ) (Form.bind y (subst_svar φ s x))

  def subst_nom (φ : Form N) (s : NOM N) (x : SVAR) : Form N :=
    match φ with
    | Form.bttm     => φ
    | Form.prop _   => φ
    | Form.svar y   => ite (x = y) s y
    | Form.nom  _   => φ
    | Form.impl φ ψ => (subst_nom φ s x) ⟶ (subst_nom ψ s x)
    | Form.box  φ   => □ (subst_nom φ s x)
    | Form.bind y φ => ite (x = y) (Form.bind y φ) (Form.bind y (subst_nom φ s x))

  def is_substable (φ : Form N) (y : SVAR) (x : SVAR) : Bool :=
    match φ with
    | Form.bttm     => true
    | Form.prop _   => true
    | Form.svar _   => true
    | Form.nom  _   => true
    | Form.impl φ ψ => (is_substable φ y x) && (is_substable ψ y x)
    | Form.box  φ   => is_substable φ y x
    | Form.bind z φ =>
        if (is_free x φ == false) then true
        else z != y && is_substable φ y x
    -- all s, s ⟶ (all x, x)  : safe,   substitution won't do anything
    -- all x, x                : safe,   substitution won't do anything
    -- all y, y ⟶ x           : safe,   result will be   all y, y ⟶ s
    -- all s, y ⟶ x           : UNSAFE, substitution would make x bound
    --                                      where it was previously free
    --
    -- Takeaway: s is substable for all free occurences of x only as long
    --         as x *does not occur free in the scope of an s-quantifier*

  notation:150 φ "[" s "//" x "]" => subst_svar φ s x
  notation:150 φ "[" s "//" x "]" => subst_nom  φ s x

end Substitutions

section NominalSubstitution

  def nom_subst_nom : Form N → NOM N → NOM N → Form N
  | .nom a, i, j     => if a = j then i else a
  | .impl φ ψ, i, j  => nom_subst_nom φ i j ⟶ nom_subst_nom ψ i j
  | .box φ, i, j     => □ nom_subst_nom φ i j
  | .bind y φ, i, j  => all y, nom_subst_nom φ i j
  | φ, _, _          => φ

  def nom_subst_svar : Form N → SVAR → NOM N → Form N
  | .nom a, i, j     => if a = j then i else a
  | .impl φ ψ, i, j  => nom_subst_svar φ i j ⟶ nom_subst_svar ψ i j
  | .box φ, i, j     => □ nom_subst_svar φ i j
  | .bind y φ, i, j  => all y, nom_subst_svar φ i j
  | φ, _, _          => φ

  notation:150 φ "[" i "//" a "]" => nom_subst_nom φ i a
  notation:150 φ "[" i "//" a "]" => nom_subst_svar φ i a

  def nom_occurs : NOM N → Form N → Bool
  | i, .nom j    => i = j
  | i, .impl ψ χ => (nom_occurs i ψ) || (nom_occurs i χ)
  | i, .box ψ    => nom_occurs i ψ
  | i, .bind _ ψ => nom_occurs i ψ
  | _, _         => false

  def all_nocc (i : NOM N) (Γ : Set (Form N)) := ∀ (φ : Form N), φ ∈ Γ → nom_occurs i φ = false

  theorem nom_occurs_conj {i : NOM N} {φ ψ : Form N} : nom_occurs i (φ ⋀ ψ) = (nom_occurs i φ || nom_occurs i ψ) := by
    show nom_occurs i ((φ ⟶ (ψ ⟶ ⊥)) ⟶ ⊥) = _
    simp only [nom_occurs, Bool.or_false]

  theorem all_noc_conj (h : all_nocc i Γ) (L : List Γ) : nom_occurs i (conjunction Γ L) = false := by
    induction L with
    | nil => simp [conjunction, nom_occurs]
    | cons head tail ih =>
        have hd : nom_occurs i head.val = false := h head head.2
        show nom_occurs i (head.val ⋀ conjunction Γ tail) = false
        rw [nom_occurs_conj, hd, ih]; rfl

  def Form.bulk_subst : Form N → List (NOM N) → List (NOM N) → Form N
  | φ, h₁ :: t₁, h₂ :: t₂ => bulk_subst (φ[h₁ // h₂]) t₁ t₂
  | φ, _, []    =>  φ
  | φ, [], _    =>  φ

  def Form.list_noms : (Form N) → List (NOM N)
  | nom  i   => [i]
  | impl φ ψ => (List.merge φ.list_noms ψ.list_noms (GE.ge · ·)).dedup
  | box φ    => φ.list_noms
  | bind _ φ => φ.list_noms
  | _        => []

  def Form.odd_list_noms : Form TotalSet → List (NOM TotalSet) := λ φ => φ.list_noms.map (λ i => 2*i+1)

  def Form.odd_noms : Form TotalSet → Form TotalSet := λ φ => φ.bulk_subst φ.odd_list_noms φ.list_noms

  def Set.odd_noms : Set (Form TotalSet) → Set (Form TotalSet) := λ Γ => {Form.odd_noms φ | φ ∈ Γ}

  def nocc_bulk_property (l1 l2 : List (NOM TotalSet)) (φ : Form TotalSet) := ∀ {n : Fin l1.length} {i : NOM TotalSet}, (i = l1[n]) → (i ∉ φ.list_noms ∨ i ∈ l2.take n) ∧ i ∉ l1.take n

  theorem list_noms_sorted_ge {φ : Form N} : φ.list_noms.Sorted GE.ge := by
    induction φ with
    | nom  i   => simp [Form.list_noms]
    | impl φ ψ ih1 ih2 =>
        exact List.Pairwise.sublist ((List.merge φ.list_noms ψ.list_noms (GE.ge · ·)).dedup_sublist) (List.Pairwise.merge ih1 ih2)
    | box _ ih    => exact ih
    | bind _ _ ih => exact ih
    | _        => simp [Form.list_noms]

  theorem list_noms_nodup {φ : Form N} : φ.list_noms.Nodup := by
    induction φ <;> simp [Form.list_noms, List.nodup_dedup, *]

  theorem list_noms_sorted_gt {φ : Form N} : φ.list_noms.Sorted GT.gt := by
    have h := List.Pairwise.and (@list_noms_sorted_ge N φ) (@list_noms_nodup N φ)
    apply List.Pairwise.imp _ h
    intro a b hab
    exact NOM.gt_iff_ge_and_ne.mpr hab

  theorem list_noms_chain' {φ : Form N} : φ.list_noms.Chain' GT.gt := by
    show List.IsChain GT.gt φ.list_noms
    rw [List.isChain_iff_pairwise]
    exact list_noms_sorted_gt

end NominalSubstitution

section IteratedModalities

  -- Axiom utils. Since we won't be assuming a transitive frame,
  -- it will make sense to be able to construct formulas with
  -- iterated modal operators at their beginning (ex., for axiom nom)
  def iterate_nec (n : Nat) (φ : Form N) : Form N :=
    let rec loop : Nat → Form  N → Form N
      | 0, φ   => φ
      | n+1, φ => □ (loop n φ)
    loop n φ

  theorem iter_nec_one : □ φ = iterate_nec 1 φ := by
    rw [iterate_nec, iterate_nec.loop, iterate_nec.loop]

  theorem iter_nec_one_m_comm : iterate_nec 1 (iterate_nec m φ) = iterate_nec m (iterate_nec 1 φ) := by
    induction m with
    | zero =>
        simp [iterate_nec, iterate_nec.loop]
    | succ n ih =>
        simp [iterate_nec, iterate_nec.loop]
        exact ih

  theorem iter_nec_compose : iterate_nec (m + 1) φ = iterate_nec m (iterate_nec 1 φ) := by
    rw [iterate_nec, iterate_nec.loop, iter_nec_one, ←iterate_nec, iter_nec_one_m_comm]

  theorem iter_nec_succ : iterate_nec (m + 1) φ = iterate_nec m (□ φ) := by
    rw [iter_nec_one, iter_nec_compose]



  def iterate_pos (n : Nat) (φ : Form N) : Form N :=
    let rec loop : Nat → Form N → Form N
      | 0, φ   => φ
      | n+1, φ => ◇ (loop n φ)
    loop n φ

  theorem iter_pos_one : ◇ φ = iterate_pos 1 φ := by
    rw [iterate_pos, iterate_pos.loop, iterate_pos.loop]

  theorem iter_pos_one_m_comm : iterate_pos 1 (iterate_pos m φ) = iterate_pos m (iterate_pos 1 φ) := by
    induction m with
    | zero =>
        simp [iterate_pos, iterate_pos.loop]
    | succ n ih =>
        simp [iterate_pos, iterate_pos.loop]
        exact ih

  theorem iter_pos_compose : iterate_pos (m + 1) φ = iterate_pos m (iterate_pos 1 φ) := by
    rw [iterate_pos, iterate_pos.loop, iter_pos_one, ←iterate_pos, iter_pos_one_m_comm]

  theorem iter_pos_succ : iterate_pos (m + 1) φ = iterate_pos m (◇ φ) := by
    rw [iter_pos_one, iter_pos_compose]


end IteratedModalities

  theorem ex_depth {x : SVAR} : Form.depth φ < Form.depth (ex x, φ) := by
    simp [Form.depth]
    rw [←Nat.add_assoc, ←Nat.add_assoc, Nat.add_comm]
\end{lstlisting}

\noindent\textcolor{green!40!black}{\textbf{Lean 4 source (lines 401--402)}}\par\vspace{0.25\baselineskip}
\begin{lstlisting}[style=leanbox,firstline=401,lastline=402]
    apply Nat.lt_add_of_pos_right
    simp
\end{lstlisting}

\hypertarget{hybridtautology.lean}{%
\subsection{Hybrid/Tautology.lean}\label{hybridtautology.lean}}

\emph{230 lines.}

\vspace{0.5\baselineskip}
\noindent\textcolor{green!40!black}{\textbf{Lean 4 source}}\par\vspace{0.25\baselineskip}
\begin{lstlisting}[style=leanbox]
import Hybrid.Form

structure Eval (N : Set ℕ) where
  f  : Form N → Bool
  p1 : f ⊥ = false
  p2 : ∀ φ ψ : Form N, (f (φ ⟶ ψ) = true) ↔ (¬(f φ) = true ∨ (f ψ) = true)

def Tautology (φ : Form N) : Prop := ∀ e : Eval N, e.f φ = true

theorem tautology_nom_subst {φ : Form N} (h : Tautology φ) (new old : NOM N) :
    Tautology (φ[new // old]) := by
  intro e
  let f : Form N → Bool := fun ψ => e.f (ψ[new // old])
  have p1 : f ⊥ = false := by simp [f, nom_subst_nom, e.p1]
  have p2 : ∀ ψ χ, f (ψ ⟶ χ) = true ↔ (¬ f ψ = true ∨ f χ = true) := by
    intro ψ χ
    show e.f ((ψ ⟶ χ)[new // old]) = true ↔ _
    simp only [nom_subst_nom, f]
    exact e.p2 (ψ[new // old]) (χ[new // old])
  exact h ⟨f, p1, p2⟩

theorem e_dn {e : Eval N} : e.f (∼φ) = false ↔ e.f φ = true := by
  rw [Form.neg, ← Bool.not_eq_true, e.p2, e.p1]
  simp [Bool.not_eq_true]

theorem e_neg {e : Eval N} : e.f (∼φ) = true ↔ e.f φ = false := by
  have c := @not_congr (e.f (∼φ) = false) (e.f φ = true) e_dn
  rw [Bool.not_eq_false, Bool.not_eq_true] at c
  exact c

theorem e_conj {e : Eval N} : e.f (φ ⋀ ψ) = true ↔ (e.f φ = true ∧ e.f ψ = true) := by
  rw [Form.conj, ←Bool.not_eq_false, e_dn, e.p2, not_or, not_not, Bool.not_eq_true, e_dn]

theorem e_disj {e : Eval N} : e.f (φ ⋁ ψ) = true ↔ (e.f φ = true ∨ e.f ψ = true) := by
  rw [Form.disj, e.p2, Bool.not_eq_true, e_dn]

theorem e_impl {e : Eval N} : e.f (φ ⟶ ψ) = true ↔ (e.f φ = true → e.f ψ = true) := by
  simp only [e.p2, implication_disjunction]

syntax "eval" : tactic
macro_rules
  | `(tactic| eval) => `(tactic| intro e; simp [e.p1, e.p2, e_dn, e_neg, e_conj, e_disj, e_impl, -Form.neg, -Form.conj, -Form.disj, -Form.iff])

theorem hs_taut : Tautology ((φ ⟶ ψ) ⟶ (ψ ⟶ χ) ⟶ (φ ⟶ χ)) := by
  intro e
  simp only [Form.iff, e_impl, e_neg, e_conj, e_disj, ← Bool.not_eq_true]
  tauto

theorem ax_1 : Tautology (φ ⟶ ψ ⟶ φ) := by
  intro e
  simp only [e.p2, Bool.not_eq_true, or_comm, ←or_assoc, Bool.dichotomy, true_or]

theorem neg_conj : Tautology ((φ ⟶ ∼ψ) ⟷ ∼(φ ⋀ ψ)) := by
  simp
  eval

theorem contrapositive : Tautology ((φ ⟶ ψ) ⟶ (∼ψ ⟶ ∼φ)) := by
  eval
  simp only [or_comm]
  simp only [and_or_right]
  apply And.intro
  . rw [←or_assoc, ←Bool.not_eq_true]
    apply Or.inl
    apply em
  . rw [←or_comm, or_assoc, or_comm, ←Bool.not_eq_true]
    apply Or.inl
    apply em

theorem contrapositive' : Tautology ((∼ψ ⟶ ∼φ) ⟶ (φ ⟶ ψ)) := by
  eval
  simp only [or_comm]
  simp only [and_or_right]
  apply And.intro
  . rw [←or_assoc, ←Bool.not_eq_true]
    apply Or.inl
    rw [or_comm]
    apply em
  . rw [←or_comm, or_assoc, or_comm, ←Bool.not_eq_true]
    apply Or.inl
    rw [or_comm]
    apply em

theorem neg_intro : Tautology ((φ ⟶ ψ) ⟶ (φ ⟶ ∼ψ) ⟶ ∼φ) := by
  intro e
  simp only [Form.iff, e_impl, e_neg, e_conj, e_disj, ← Bool.not_eq_true]
  tauto

theorem imp_refl : Tautology (φ ⟶ φ) := by
  eval

theorem imp_neg : Tautology (∼(φ ⟶ ψ) ⟷ (φ ⋀ ∼ψ)) := by
  simp only [Form.iff, Form.conj, Form.neg]
  eval

theorem dne : Tautology (∼∼φ ⟶ φ) := by
  eval

theorem dni : Tautology (φ ⟶ ∼∼φ) := by
  eval

theorem dn : Tautology (φ ⟷ ∼∼φ) := by
  intro e
  rw [Form.iff, e_conj]
  exact ⟨dni e, dne e⟩ 

theorem conj_intro : Tautology (φ ⟶ ψ ⟶ (φ ⋀ ψ)) := by
  intro e
  simp only [Form.iff, e_impl, e_neg, e_conj, e_disj, ← Bool.not_eq_true]
  tauto

theorem conj_intro_hs : Tautology ((φ ⟶ ψ) ⟶ (φ ⟶ χ) ⟶ (φ ⟶ (ψ ⋀ χ))) := by
  intro e
  simp only [Form.iff, e_impl, e_neg, e_conj, e_disj, ← Bool.not_eq_true]
  tauto

theorem conj_elim_l : Tautology ((φ ⋀ ψ) ⟶ φ) := by
  eval
  simp [←or_assoc, or_comm, Bool.dichotomy]

theorem conj_elim_r : Tautology ((φ ⋀ ψ) ⟶ ψ) := by
  eval
  simp [or_assoc, Bool.dichotomy]

theorem conj_comm_t : Tautology ((φ ⋀ ψ) ⟶ (ψ ⋀ φ)) := by
  intro e
  simp only [e_impl, e_conj]
  tauto

theorem conj_comm_t' : Tautology (∼(φ ⋀ ψ) ⟶ ∼(ψ ⋀ φ)) := by
  intro e
  simp only [e_impl, e_neg, ← Bool.not_eq_true, e_conj]
  tauto

theorem iff_intro : Tautology ((φ ⟶ ψ) ⟶ (ψ ⟶ φ) ⟶ (φ ⟷ ψ)) := by
  intro e
  simp only [Form.iff, e_impl, e_neg, e_conj, e_disj, ← Bool.not_eq_true]
  tauto

theorem iff_elim_l : Tautology ((φ ⟷ ψ) ⟶ (φ ⟶ ψ)) := by
  intro e
  simp only [Form.iff, e_impl, e_neg, e_conj, e_disj, ← Bool.not_eq_true]
  tauto

theorem iff_elim_r : Tautology ((φ ⟷ ψ) ⟶ (ψ ⟶ φ)) := by
  intro e
  simp only [Form.iff, e_impl, e_neg, e_conj, e_disj, ← Bool.not_eq_true]
  tauto

theorem iff_rw : Tautology ((φ ⟷ ψ) ⟶ (ψ ⟷ χ) ⟶ (φ ⟷ χ)) := by
  intro e
  simp only [Form.iff, e_impl, e_neg, e_conj, e_disj, ← Bool.not_eq_true]
  tauto

theorem iff_imp : Tautology ((φ ⟷ ψ) ⟶ (χ ⟷ τ) ⟶ ((φ ⟶ χ) ⟷ (ψ ⟶ τ))) := by
  intro e
  simp only [Form.iff, e_impl, e_neg, e_conj, e_disj, ← Bool.not_eq_true]
  tauto

theorem taut_iff_mp : Tautology (φ ⟷ ψ) → Tautology (φ ⟶ ψ) := by
  rw [Form.iff]
  intro h e
  have := h e
  rw [e_conj] at this
  exact this.left

theorem taut_iff_mpr : Tautology (φ ⟷ ψ) → Tautology (ψ ⟶ φ) := by
  rw [Form.iff]
  intro h e
  have := h e
  rw [e_conj] at this
  exact this.right

theorem disj_intro_l : Tautology (φ ⟶ (φ ⋁ ψ)) := by
  intro e
  simp only [Form.iff, e_impl, e_neg, e_conj, e_disj, ← Bool.not_eq_true]
  tauto

theorem disj_intro_r : Tautology (φ ⟶ (ψ ⋁ φ)) := by
  intro e
  simp only [Form.iff, e_impl, e_neg, e_conj, e_disj, ← Bool.not_eq_true]
  tauto

theorem disj_elim : Tautology ((φ ⋁ ψ) ⟶ (φ ⟶ χ) ⟶ (ψ ⟶ χ) ⟶ χ) := by
  intro e
  simp only [Form.iff, e_impl, e_neg, e_conj, e_disj, ← Bool.not_eq_true]
  tauto

theorem idem : Tautology ((χ ⟶ ψ ⟶ ψ ⟶ φ) ⟶ (χ ⟶ ψ ⟶ φ)) := by
  intro e
  simp only [e_impl]
  tauto

theorem exp : Tautology (((φ ⋀ ψ) ⟶ χ) ⟶ (φ ⟶ ψ ⟶ χ)) := by
  intro e
  simp only [e_impl, e_conj]
  tauto

theorem imp : Tautology ((φ ⟶ ψ ⟶ χ) ⟶ ((φ ⋀ ψ)) ⟶ χ) := by
  intro e
  simp only [e_impl, e_conj]
  tauto

theorem impexp : Tautology (((φ ⋀ ψ) ⟶ χ) ⟷ (φ ⟶ ψ ⟶ χ)) := by
  intro e
  rw [Form.iff, e_conj]
  exact ⟨exp e, imp e⟩

theorem com12 : Tautology ((φ ⟶ (ψ ⟶ χ)) ⟶ (ψ ⟶ (φ ⟶ χ))) := by
  intro e
  simp only [e_impl]
  intro h1 h2 h3
  exact h1 h3 h2

theorem mp_help : Tautology ((a ⟶ (φ ⟶ ψ)) ⟶ ((b ⟶ φ) ⟶ (a ⟶ b ⟶ ψ))) := by
  intro e
  simp only [Form.iff, e_impl, e_neg, e_conj, e_disj, ← Bool.not_eq_true]
  tauto

def Eval.nom_variant (e e' : Eval N) (i : NOM N) (x : SVAR) : Prop :=
  e'.f = (λ φ : Form N => if φ = i then (e.f x) else (e.f φ))

theorem iff_not : Tautology ((φ ⟷ ψ) ⟷ (∼φ ⟷ ∼ψ)) := by
  simp only [Form.iff, Form.conj, Form.neg]
  eval
 
theorem imp_taut (h : Tautology φ) : Tautology ((φ ⟶ ψ) ⟶ ψ) := by
  unfold Tautology at h ⊢
  intro e
  have := h e
  simp [this, e.p1, e.p2, e_dn, e_neg, e_conj, e_disj, e_impl, -Form.neg, -Form.conj, -Form.disj, -Form.iff]
\end{lstlisting}
\vspace{0.5\baselineskip}

\hypertarget{hybridsubstitutions.lean}{%
\subsection{Hybrid/Substitutions.lean}\label{hybridsubstitutions.lean}}

\emph{1228 lines.}

\noindent\textcolor{green!40!black}{\textbf{Lean 4 source (lines 1--400)}}\par\vspace{0.25\baselineskip}
\begin{lstlisting}[style=leanbox,firstline=1,lastline=400]
import Hybrid.Form

-- Helper simp lemmas to reduce SVAR ordering / max / addition to `Nat`,
-- so that `omega` can finish arithmetic goals. (mathlib upgrade: the old
-- `simp [SVAR.le, max, SVAR.add]` patterns no longer unfold reliably.)
theorem svar_le_letter {x y : SVAR} : (x ≤ y) = (x.letter ≤ y.letter) := rfl
theorem svar_lt_letter {x y : SVAR} : (x < y) = (x.letter < y.letter) := rfl
theorem svar_add_letter {x : SVAR} {n : Nat} : (x + n).letter = x.letter + n := rfl
theorem svar_max_letter {x y : SVAR} : (max x y).letter = max x.letter y.letter := by
  change (ite (x.letter > y.letter) x y).letter = max x.letter y.letter
  split <;> omega

theorem subst_depth {i : NOM N} {x : SVAR} {φ : Form N} : φ[i // x].depth = φ.depth := by
  induction φ <;> simp [subst_nom, Form.depth, *] at *
  <;> (split <;> simp [Form.depth, *])

theorem subst_depth' {x y : SVAR} {φ : Form N} : φ[y // x].depth = φ.depth := by
  induction φ <;> simp [subst_svar, Form.depth, *] at *
  <;> (split <;> simp [Form.depth, *])

theorem subst_depth'' {x : SVAR} {i : NOM N} {φ : Form N} : (φ[i//x]).depth < (ex x, φ).depth := by
  apply Nat.lt_of_le_of_lt
  apply Nat.le_of_eq
  apply subst_depth
  apply ex_depth

theorem subst_depth_bind {x : SVAR} {i : NOM N} {φ : Form N} : (φ[i//x]).depth < (all x, φ).depth := by
  apply Nat.lt_of_le_of_lt
  apply Nat.le_of_eq
  apply subst_depth
  apply sub_depth_bind

theorem iff_subst_svar {y x : SVAR} : (φ ⟷ ψ)[y // x] = (φ[y//x] ⟷ ψ[y//x]) := by
  simp [subst_svar]

section Variables
  lemma svar_eq {ψ χ : SVAR} : ψ = χ ↔ ψ.1 = χ.1 := by
    have l1 : ψ = ⟨ψ.letter⟩ := by simp
    have l2 : χ = ⟨χ.letter⟩ := by simp
    rw [l1, l2]
    simp

  lemma new_var_neg : (∼ψ).new_var = ψ.new_var := by
    simp [Form.new_var, max, -implication_disjunction]
    rw [←svar_eq]
    intro _
    simp [*]
  
  lemma subst_neg : is_substable (∼ψ) y x ↔ is_substable ψ y x := by
    simp [is_substable]

  lemma new_var_gt      : occurs x φ → x < φ.new_var   := by
    induction φ with
    | svar y          =>
        simp [occurs, Form.new_var, -implication_disjunction]
        intro h
        rw [h]
        exact Nat.lt_succ_self y.letter
    | impl ψ χ ih1 ih2 =>
        simp only [occurs, Form.new_var, Bool.or_eq_true, max]
        intro h
        apply Or.elim h
        . intro ha
          clear ih2 h
          have ih1 := ih1 ha
          by_cases hc : (Form.new_var ψ).letter > (Form.new_var χ).letter
          . simp [hc]
            assumption
          . simp [hc]
            simp at hc 
            exact Nat.lt_of_lt_of_le ih1 hc
        . intro hb
          clear ih1 h
          have ih2 := ih2 hb
          by_cases hc : (Form.new_var ψ).letter > (Form.new_var χ).letter
          . simp [hc]
            simp at hc
            exact Nat.lt_trans ih2 hc
          . simp [hc]
            assumption
    | box ψ ih      =>
        simp only [occurs, Form.new_var]
        assumption
    | bind y ψ ih   =>
        simp only [occurs, Form.new_var, max]
        intro h
        have ih := ih h
        by_cases hc : (y + 1).letter > (Form.new_var ψ).letter
        . simp [hc]
          simp at hc
          exact Nat.lt_trans ih hc
        . simp [hc]
          assumption
    | _ => simp [occurs]

  lemma new_var_is_new  : occurs (φ.new_var) φ = false := by
    rw [←Bool.eq_false_eq_not_eq_true]
    intro h
    have a := new_var_gt h
    have b := Nat.lt_irrefl φ.new_var.letter
    exact b a
  
  lemma ge_new_var_is_new (h : x ≥ φ.new_var) : occurs x φ = false := by
    rw [←Bool.eq_false_eq_not_eq_true]
    intro habs
    have := new_var_gt habs
    have a := Nat.lt_of_le_of_lt h this
    have b := Nat.lt_irrefl φ.new_var.letter
    exact b a
  
  lemma ge_new_var_subst_nom {i : NOM N} {y : SVAR} : φ.new_var ≥ φ[i // y].new_var := by
    induction φ with
    | svar z =>
        simp only [subst_nom]
        split <;> simp [Form.new_var, svar_le_letter, svar_add_letter]
    | impl ψ χ ih1 ih2 =>
        simp only [subst_nom, Form.new_var, ge_iff_le, svar_le_letter, svar_max_letter] at *
        omega
    | bind z ψ ih =>
        simp only [subst_nom]
        split <;>
          simp only [Form.new_var, ge_iff_le, svar_le_letter, svar_max_letter, svar_add_letter] at * <;>
          omega
    | box ψ ih =>
        simpa only [subst_nom, Form.new_var, ge_iff_le, svar_le_letter] using ih
    | _ => simp [Form.new_var, subst_nom, svar_le_letter]

lemma new_var_geq1 : x ≥ (φ ⟶ ψ).new_var → (x ≥ φ.new_var ∧ x ≥ ψ.new_var) := by
  intro h
  simp [Form.new_var, max] at *
  split at h
  . apply And.intro
    . assumption
    . apply Nat.le_trans _ h
      apply Nat.le_of_lt
      assumption
  . apply And.intro
    . simp at *
      apply Nat.le_trans _ h
      assumption
    . assumption

lemma new_var_geq2 : x ≥ (all y, ψ).new_var → (x ≥ (y+1) ∧ x ≥ ψ.new_var) := by
  intro h
  simp [Form.new_var, max] at *
  split at h
  . apply And.intro
    . apply Nat.le_trans _ h
      apply Nat.le_of_lt
      assumption
    . assumption
  . apply And.intro
    . assumption
    . simp at *
      apply Nat.le_trans _ h
      assumption

lemma new_var_geq3 : x ≥ (□ φ).new_var → (x ≥ φ.new_var) := by simp [Form.new_var]

lemma new_var_subst {φ : Form N} {i : NOM N} {x y : SVAR} (h : x ≥ φ.new_var) : is_substable (φ[y//i]) x y := by
  induction φ with
  | nom  j  =>
      simp [nom_subst_svar]
      split <;> simp [is_substable]
  | bind z ψ ih =>
      simp only [nom_subst_svar, Form.new_var, max, is_substable, beq_iff_eq, ite_eq_left_iff,
          bne, Bool.not_eq_true', beq_eq_false_iff_ne, ne_eq,
          Bool.not_eq_false, Bool.and_eq_true] at h ⊢
      intro _
      by_cases hc : (z + 1).letter > (Form.new_var ψ).letter
      . simp [hc] at h
        simp only [gt_iff_lt, ge_iff_le] at hc ih
        have ih := ih (Nat.le_of_lt (Nat.lt_of_lt_of_le hc h))
        have ne := Nat.ne_of_lt (Nat.lt_of_lt_of_le (Nat.lt_succ_self z.letter) h)
        rw [of_eq_true (eq_self z), of_eq_true (eq_self x), SVAR.mk.injEq]
        exact ⟨ne, ih⟩
      . simp [hc] at h
        simp only [gt_iff_lt, not_lt, ge_iff_le] at hc ih 
        have ih := ih h
        have ne := Nat.ne_of_lt (Nat.le_trans (Nat.lt_of_lt_of_le (Nat.lt_succ_self z.letter) hc) h)
        rw [of_eq_true (eq_self z), of_eq_true (eq_self x), SVAR.mk.injEq]
        exact ⟨ne, ih⟩
  | impl ψ χ ih1 ih2 =>
      simp [Form.new_var, max, is_substable, nom_subst_svar] at h ⊢
      by_cases hc : (Form.new_var χ).letter < (Form.new_var ψ).letter
      . simp [hc] at h
        have := Nat.le_of_lt (Nat.lt_of_lt_of_le hc h)
        exact ⟨ih1 h, ih2 this⟩
      . simp [hc] at h
        simp at hc
        have := Nat.le_trans hc h
        exact ⟨ih1 this, ih2 h⟩
  | box ψ ih         =>
      simp [Form.new_var, is_substable, nom_subst_svar] at h ⊢
      exact ih h
  | _  =>
      simp [nom_subst_svar, is_substable]

lemma new_var_subst'' {φ : Form N} {x y : SVAR} (h : x ≥ φ.new_var) : is_substable φ x y := by
  induction φ with
  | bind z ψ ih =>
      simp only [Form.new_var, max, is_substable, beq_iff_eq, ite_eq_left_iff,
          bne, Bool.not_eq_true', beq_eq_false_iff_ne, ne_eq,
          Bool.not_eq_false, Bool.and_eq_true] at h ⊢
      intro _
      by_cases hc : (z + 1).letter > (Form.new_var ψ).letter
      . simp [hc] at h
        simp only [gt_iff_lt, ge_iff_le] at hc ih
        have ih := ih (Nat.le_of_lt (Nat.lt_of_lt_of_le hc h))
        have ne := Nat.ne_of_lt (Nat.lt_of_lt_of_le (Nat.lt_succ_self z.letter) h)
        rw [of_eq_true (eq_self z), of_eq_true (eq_self x), SVAR.mk.injEq]
        exact ⟨ne, ih⟩
      . simp [hc] at h
        simp only [gt_iff_lt, not_lt, ge_iff_le] at hc ih 
        have ih := ih h
        have ne := Nat.ne_of_lt (Nat.le_trans (Nat.lt_of_lt_of_le (Nat.lt_succ_self z.letter) hc) h)
        rw [of_eq_true (eq_self z), of_eq_true (eq_self x), SVAR.mk.injEq]
        exact ⟨ne, ih⟩
  | impl ψ χ ih1 ih2 =>
      simp [Form.new_var, max, is_substable, nom_subst_svar] at h ⊢
      by_cases hc : (Form.new_var χ).letter < (Form.new_var ψ).letter
      . simp [hc] at h
        have := Nat.le_of_lt (Nat.lt_of_lt_of_le hc h)
        exact ⟨ih1 h, ih2 this⟩
      . simp [hc] at h
        simp at hc
        have := Nat.le_trans hc h
        exact ⟨ih1 this, ih2 h⟩
  | box ψ ih         =>
      simp [Form.new_var, is_substable, nom_subst_svar] at h ⊢
      exact ih h
  | _  =>
      simp [is_substable]

lemma scz {φ : Form N} (i : NOM N) (h : x ≥ φ.new_var) (hy : y ≠ x) : (is_free y φ) ↔ (is_free y (φ[x // i])) := by
  induction φ with
  | nom a       =>
      simp [nom_subst_svar] ; split <;> simp [is_free, hy]
  | bind z ψ ih =>
      simp [is_free, nom_subst_svar, -implication_disjunction]
      simp [new_var_geq2 h] at ih
      simp [nom_subst_svar, is_free, ih]
  | impl ψ χ ih1 ih2 =>
      have ⟨ih1_cond, ih2_cond⟩ := new_var_geq1 h
      simp [ih1_cond, ih2_cond] at ih1 ih2
      simp [is_free, nom_subst_svar, ih1, ih2]
  | box ψ ih         =>
      simp [Form.new_var] at h
      simp [h] at ih
      simp [is_free, nom_subst_svar, ih]
  | _ => simp [is_free, nom_subst_svar]

lemma new_var_subst' {φ : Form N} (i : NOM N) {x y : SVAR} (h1 : is_substable φ v y) (h2 : x ≥ φ.new_var) (h3 : y ≠ x) : is_substable (φ[x//i]) v y := by
  induction φ with
  | nom  a      => simp [nom_subst_svar]; split <;> simp [is_substable]
  | bind z ψ ih =>
      have xge := (new_var_geq2 h2).right
      have hsc := @scz N x y ψ i xge h3
      have heq : is_free y (ψ[x//i]) = is_free y ψ := by
        cases h' : is_free y ψ <;> cases h'' : is_free y (ψ[x//i]) <;> simp_all
      simp only [nom_subst_svar]
      simp [is_substable] at h1 ⊢
      rcases h1 with hf | ⟨hzv, hsub⟩
      · left; rw [heq]; exact hf
      · right; exact ⟨hzv, ih hsub xge⟩
  | impl ψ χ ih1 ih2  =>
      simp [is_substable] at h1
      simp [Form.new_var] at h2
      have ⟨ih1_cond, ih2_cond⟩ := new_var_geq1 h2 
      simp [h1, h2, ih1_cond, ih2_cond] at ih1 ih2
      simp [is_substable, nom_subst_svar, ih1, ih2]
  | box ψ ih          =>
      simp [is_substable] at h1
      simp [Form.new_var] at h2
      simp [h1, h2] at ih
      simp [is_substable, nom_subst_svar, ih]
  | _       =>  simp [nom_subst_svar, h1]

lemma nom_subst_trans (i : NOM N) (x y : SVAR) (h : y ≥ φ.new_var) : φ[y // i][x // y] = φ[x // i] := by
  induction φ with
  | bttm => simp [nom_subst_svar, subst_svar]
  | prop => simp [nom_subst_svar, subst_svar]
  | nom _ =>
    simp [nom_subst_svar]
    split <;> simp [subst_svar]
  | svar z =>
    have nocc := ge_new_var_is_new h
    simp only [nom_subst_svar, subst_svar]
    split <;> simp_all [occurs]
  | bind z ψ ih =>
    simp only [nom_subst_svar, subst_svar]
    have := new_var_geq2 h
    by_cases hc : y = z
    . exfalso
      have := this.left
      simp [hc] at this
      have := Nat.ne_of_lt (Nat.lt_succ_of_le this)
      contradiction
    . simp [nom_subst_svar, ih this.right, hc]
  | impl ψ χ ih1 ih2 =>
      simp [nom_subst_svar, subst_svar, ih1, ih2, new_var_geq1 h]
  | box ψ ih         =>
      simp [Form.new_var] at h
      simp [nom_subst_svar, subst_svar, ih, h]

  lemma subst_nom_noop {φ : Form N} {i : NOM N} {y : SVAR} (h : occurs y φ = false) : φ[i // y] = φ := by
    induction φ with
    | svar z => simp only [subst_nom]; split <;> simp_all [occurs]
    | impl a b iha ihb =>
        simp only [occurs, Bool.or_eq_false_iff] at h
        simp only [subst_nom, iha h.1, ihb h.2]
    | box a ih => simp only [occurs] at h; simp only [subst_nom, ih h]
    | bind w a ih => simp only [occurs] at h; simp only [subst_nom]; split <;> simp_all [ih h]
    | _ => rfl

  -- Substituting `x` by a fresh `y` and then renaming `y` to a nominal `i` is the
  -- same as directly substituting `x` by `i`.  Freshness of `y` (it exceeds the
  -- new-variable bound, so it differs from every free *and bound* variable of `φ`)
  -- is what prevents capture.
  lemma rename_svar_nom {φ : Form N} (i : NOM N) (x y : SVAR) (h : y ≥ φ.new_var) : φ[y // x][i // y] = φ[i // x] := by
    induction φ with
    | bttm => simp [subst_svar, subst_nom]
    | prop => simp [subst_svar, subst_nom]
    | nom _ => simp [subst_svar, subst_nom]
    | svar z =>
        have hyz : y ≠ z := by
          have := ge_new_var_is_new h; simpa [occurs] using this
        by_cases hxz : x = z
        · subst hxz; simp [subst_svar, subst_nom]
        · simp [subst_svar, subst_nom, hxz, hyz]
    | impl ψ χ ih1 ih2 =>
        simp [subst_svar, subst_nom, ih1 (new_var_geq1 h).1, ih2 (new_var_geq1 h).2]
    | box ψ ih =>
        simp only [Form.new_var] at h
        simp [subst_svar, subst_nom, ih h]
    | bind z ψ ih =>
        have hb := new_var_geq2 h
        have hyz : y ≠ z := by
          intro habs; have hle := hb.left; rw [habs] at hle
          simp only [svar_le_letter, svar_add_letter] at hle; omega
        by_cases hxz : x = z
        · subst hxz
          have hnoop : ψ[i // y] = ψ := subst_nom_noop (ge_new_var_is_new hb.right)
          simp [subst_svar, subst_nom, hyz, hnoop]
        · simp [subst_svar, subst_nom, hxz, hyz, ih hb.right]

  lemma ge_new_var_subst_helpr {i : NOM N} {x : SVAR} (h : y ≥ Form.new_var (χ⟶ψ)) : y ≥ Form.new_var (χ⟶ψ[i//x]⟶⊥) := by
    simp [Form.new_var, max]
    split <;> split
    . exact (new_var_geq1 h).left
    . apply Nat.le_trans
      apply ge_new_var_subst_nom
      exact (new_var_geq1 h).right
    . exact (new_var_geq1 h).left
    . simp [svar_le_letter]

  lemma notfreeset {Γ : Set (Form N)} (L : List Γ) (hyp : ∀ ψ : Γ, is_free x ψ.1 = false) : is_free x (conjunction Γ L) = false := by
    induction L with
    | nil         =>
        simp [conjunction, is_free]
    | cons hd tl ih =>
        have hhd := hyp hd
        simp [conjunction, is_free, hhd, ih]

  lemma notfree_after_subst {φ : Form N} {x y : SVAR} (h : x ≠ y) : is_free x (φ[y // x]) = false := by
    induction φ with
    | svar z   =>
        by_cases xz : x = z
        . simp [subst_svar, if_pos xz, is_free, h]
        . simp [subst_svar, if_neg xz, is_free, xz]
    | impl _ _ ih1 ih2 =>
        simp [subst_svar, is_free, ih1, ih2]
    | box _ ih    =>
        simp [subst_svar, is_free, ih]
    | bind z _ ih =>
        by_cases xz : x = z
        . simp [subst_svar, xz, is_free]
        . simp [subst_svar, if_neg xz, is_free, ih]
    | _        => simp [subst_svar, is_free]

  lemma notocc_beforeafter_subst {φ : Form N} {x y : SVAR} (h : occurs x φ = false) : occurs x (φ[y // x]) = false := by
    induction φ with
    | svar z   =>
        by_cases xz : x = z
        <;> simp [subst_svar, if_pos xz, xz, occurs, h] at *
    | impl _ _ ih1 ih2 =>
        simp [subst_svar, occurs, not_or, ih1, ih2, -implication_disjunction] at *
        exact ⟨ih1 h.left, ih2 h.right⟩ 
    | box _ ih    =>
        simp [subst_svar, occurs, ih, -implication_disjunction] at *
        exact ih h
    | bind z ψ ih =>
        by_cases xz : x = z
        . simp [subst_svar, xz, occurs] at *
          exact h
        . simp [subst_svar, if_neg xz, occurs, ih, xz, h, -implication_disjunction] at *
    | _        => simp [subst_svar, occurs]

  lemma notoccursbind : occurs x φ = false → occurs x (all v, φ) = false := by
    simp [occurs]
\end{lstlisting}

\noindent\textcolor{green!40!black}{\textbf{Lean 4 source (lines 401--800)}}\par\vspace{0.25\baselineskip}
\lstinputlisting[style=leanbox,firstline=401,lastline=800]{lean-listings/Hybrid-Substitutions.lean}

\noindent\textcolor{green!40!black}{\textbf{Lean 4 source (lines 801--1200)}}\par\vspace{0.25\baselineskip}
\lstinputlisting[style=leanbox,firstline=801,lastline=1200]{lean-listings/Hybrid-Substitutions.lean}

\noindent\textcolor{green!40!black}{\textbf{Lean 4 source (lines 1201--1228)}}\par\vspace{0.25\baselineskip}
\begin{lstlisting}[style=leanbox,firstline=1201,lastline=1228]
    exact bulk_subst_bind

  theorem odd_conj_two {φ ψ : Form TotalSet} : (φ ⋀ ψ).odd_noms = φ.odd_noms ⋀ ψ.odd_noms := by
    simp only [Form.conj, Form.neg, odd_impl, odd_bttm]

  def List.to_odd {Γ : Set (Form TotalSet)} : List Γ → List Γ.odd_noms
    | [] => []
    | (h :: t) => ⟨h.val.odd_noms, ⟨h.val, h.2, rfl⟩⟩ :: t.to_odd

  noncomputable def List.odd_to {Γ : Set (Form TotalSet)} : List Γ.odd_noms → List Γ
    | [] => []
    | (h :: t) => ⟨h.2.choose, h.2.choose_spec.1⟩ :: t.odd_to

  theorem odd_conj (Γ : Set (Form TotalSet)) (L : List Γ) : (conjunction Γ L).odd_noms = conjunction Γ.odd_noms L.to_odd := by
    induction L with
    | nil => simp only [conjunction, List.to_odd, odd_impl, odd_bttm]
    | cons head tail ih =>
        show (head.val ⋀ conjunction Γ tail).odd_noms
            = head.val.odd_noms ⋀ conjunction Γ.odd_noms tail.to_odd
        rw [odd_conj_two, ih]

  theorem odd_conj_rev (Γ : Set (Form TotalSet)) (L' : List Γ.odd_noms) : (conjunction Γ L'.odd_to).odd_noms = conjunction Γ.odd_noms L' := by
    induction L' with
    | nil => simp only [conjunction, List.odd_to, odd_impl, odd_bttm]
    | cons head tail ih =>
        show (head.2.choose ⋀ conjunction Γ tail.odd_to).odd_noms
            = head.val ⋀ conjunction Γ.odd_noms tail
        rw [odd_conj_two, ih, head.2.choose_spec.2]
\end{lstlisting}

\hypertarget{hybridproof.lean}{%
\subsection{Hybrid/Proof.lean}\label{hybridproof.lean}}

\emph{111 lines.}

\vspace{0.5\baselineskip}
\noindent\textcolor{green!40!black}{\textbf{Lean 4 source}}\par\vspace{0.25\baselineskip}
\begin{lstlisting}[style=leanbox]
import Hybrid.Form
import Hybrid.Tautology

inductive Proof : Form N → Type where
  -- Deduction rules:

  -- if φ is a theorem, ∀ v, φ is a theorem
  | general {φ : Form N} (v : SVAR):
        Proof φ → Proof (all v, φ)

  -- if φ is a theorem, □ φ is a theorem
  | necess {φ : Form N}:
        Proof φ → Proof (□ φ)

  -- modus ponens:
  | mp {φ ψ : Form N}:
        Proof (φ ⟶ ψ) → Proof φ → Proof ψ

  -- All propositional tautologies
  | tautology {φ : Form N}:
        Tautology φ → Proof φ

  -- Axioms for modal + hybrid logic:
  -- distribution schema (axiom K)
  | ax_k {φ ψ : Form N}:
        Proof (□ (φ ⟶ ψ) ⟶ (□ φ ⟶ □ ψ))

  | ax_q1 (φ ψ : Form N) {v : SVAR} (p : is_free v φ = false):
        Proof ((all v, φ ⟶ ψ) ⟶ (φ ⟶ all v, ψ))

  -- two different instances of Axiom Q2: one for SVAR, one for NOM
  | ax_q2_svar (φ : Form N) (v s : SVAR) (p : is_substable φ s v):
      Proof ((all v, φ) ⟶ φ[s // v])

  | ax_q2_nom (φ : Form N) (v : SVAR) (s : NOM N):
      Proof ((all v, φ) ⟶ φ[s // v])

  | ax_name (v : SVAR):
      Proof (ex v, v)

  | ax_nom {φ : Form N} {v : SVAR} (m n : Nat):
      Proof (all v, (iterate_pos m (v ⋀ φ) ⟶ iterate_nec n (v ⟶ φ)))

  | ax_brcn {φ : Form N} {v : SVAR}:
      Proof ((all v, □ φ) ⟶ (□ all v, φ))

def Proof.size : Proof φ → ℕ
  | .general _ pf => pf.size + 1
  | .necess pf    => pf.size + 1
  | .mp pf1 pf2   => pf1.size + pf2.size + 1
  | _ => 1

lemma Proof.size_lt_mp {α β : Form N} (pf1 : Proof (α ⟶ β)) (pf2 : Proof α) :
    pf1.size < (mp pf1 pf2).size := by simp [Proof.size]; omega

lemma Proof.size_lt_mp₂ {α β : Form N} (pf1 : Proof (α ⟶ β)) (pf2 : Proof α) :
    pf2.size < (mp pf1 pf2).size := by simp [Proof.size]; omega

lemma Proof.size_lt_general {φ : Form N} (v : SVAR) (pf : Proof φ) :
    pf.size < (general v pf).size := by simp [Proof.size]

lemma Proof.size_lt_necess {φ : Form N} (pf : Proof φ) :
    pf.size < (necess pf).size := by simp [Proof.size]

def Proof.contains {φ : Form N} : Proof φ → Form N → Bool :=
  λ pf ψ => φ == ψ ||
    match pf with
    | .general _ pf' => pf'.contains φ
    | .necess pf'    => pf'.contains φ
    | .mp pf1 pf2    => pf1.contains φ || pf2.contains φ
    | _ => false

def Proof.fresh_var : Proof φ → SVAR → Prop := λ pf x => ∀ ψ, pf.contains ψ → x ≥ ψ.new_var

/-- Every formula root appearing in a derivation (for nominal inventory / conservativity). -/
def Proof.formulasIn {φ : Form N} : Proof φ → List (Form N)
  | .tautology _ => [φ]
  | .ax_k => [φ]
  | .ax_q1 _ _ _ => [φ]
  | .ax_q2_svar _ _ _ _ => [φ]
  | .ax_q2_nom _ _ _ => [φ]
  | .ax_name _ => [φ]
  | .ax_nom _ _ => [φ]
  | .ax_brcn => [φ]
  | .general _ pf => φ :: pf.formulasIn
  | .necess pf => φ :: pf.formulasIn
  | .mp pf1 pf2 => φ :: pf1.formulasIn ++ pf2.formulasIn

/-- Nominals occurring in any formula of a derivation (deduped, descending merge order). -/
def Proof.proof_noms {φ : Form N} (pf : Proof φ) : List (NOM N) :=
  (pf.formulasIn.flatMap Form.list_noms).dedup

def SyntacticConsequence (Γ : Set (Form N)) (φ : Form N) := Σ L, Proof ((conjunction Γ L) ⟶ φ)

prefix:500 "⊢"  => Proof
infix:500 "⊢"   => SyntacticConsequence

notation "⊬" φ    => (Proof φ) → False
notation Γ "⊬" φ  => (SyntacticConsequence Γ φ) → False


def consistent (Γ : Set (Form N)) := ∀ (_ : SyntacticConsequence Γ ⊥), False

def MCS (Γ : Set (Form N)) := consistent Γ ∧ (∀ {φ : Form N}, (¬φ ∈ Γ) → ¬consistent (Γ ∪ {φ}))

def witnessed (Γ : Set (Form N)) : Prop := ∀ {φ : Form N},
  φ ∈ Γ →
    match φ with
--      | ex x, ψ => ∃ i : NOM, ((ex x, ψ) ⟶ ψ[i // x]) ∈ Γ
      | ex x, ψ => ∃ i : NOM N, ψ[i // x] ∈ Γ
      | _   => φ ∈ Γ
\end{lstlisting}
\vspace{0.5\baselineskip}

\hypertarget{hybridtruth.lean}{%
\subsection{Hybrid/Truth.lean}\label{hybridtruth.lean}}

\emph{331 lines.}

\vspace{0.5\baselineskip}
\noindent\textcolor{green!40!black}{\textbf{Lean 4 source}}\par\vspace{0.25\baselineskip}
\lstinputlisting[style=leanbox]{lean-listings/Hybrid-Truth.lean}
\vspace{0.5\baselineskip}

\hypertarget{hybridlistutils.lean}{%
\subsection{Hybrid/ListUtils.lean}\label{hybridlistutils.lean}}

\emph{212 lines.}

\vspace{0.5\baselineskip}
\noindent\textcolor{green!40!black}{\textbf{Lean 4 source}}\par\vspace{0.25\baselineskip}
\lstinputlisting[style=leanbox]{lean-listings/Hybrid-ListUtils.lean}
\vspace{0.5\baselineskip}

\hypertarget{hybridproofutils.lean}{%
\subsection{Hybrid/ProofUtils.lean}\label{hybridproofutils.lean}}

\emph{961 lines.}

\noindent\textcolor{green!40!black}{\textbf{Lean 4 source (lines 1--400)}}\par\vspace{0.25\baselineskip}
\lstinputlisting[style=leanbox,firstline=1,lastline=400]{lean-listings/Hybrid-ProofUtils.lean}

\noindent\textcolor{green!40!black}{\textbf{Lean 4 source (lines 401--800)}}\par\vspace{0.25\baselineskip}
\lstinputlisting[style=leanbox,firstline=401,lastline=800]{lean-listings/Hybrid-ProofUtils.lean}

\noindent\textcolor{green!40!black}{\textbf{Lean 4 source (lines 801--961)}}\par\vspace{0.25\baselineskip}
\lstinputlisting[style=leanbox,firstline=801,lastline=961]{lean-listings/Hybrid-ProofUtils.lean}

\hypertarget{hybridsoundness.lean}{%
\subsection{Hybrid/Soundness.lean}\label{hybridsoundness.lean}}

\emph{421 lines.}

\noindent\textcolor{green!40!black}{\textbf{Lean 4 source (lines 1--400)}}\par\vspace{0.25\baselineskip}
\lstinputlisting[style=leanbox,firstline=1,lastline=400]{lean-listings/Hybrid-Soundness.lean}

\noindent\textcolor{green!40!black}{\textbf{Lean 4 source (lines 401--421)}}\par\vspace{0.25\baselineskip}
\begin{lstlisting}[style=leanbox,firstline=401,lastline=421]
    assumption

theorem Consistency : ∀ {N : Set ℕ}, ⊬ (@Form.bttm N) := by
  intro N habs
  have bot_valid := WeakSoundness habs
  let M : Model N := ⟨ℕ, λ _ => λ _ => True, λ _ => ∅,  λ _ => 0⟩
  have g : I (M.W) := λ _ => 0
  have := bot_valid M 0 g
  exact this

theorem npf_negpf : ⊢ (∼φ) → ⊬ φ := by
  intro h habs
  have := Proof.mp h habs
  exact Consistency this

theorem pos_npf_not : ⊢(◇φ) → ⊬(∼φ) := by
  rw [Form.diamond]
  intro h habs
  have := Proof.necess habs
  have := Proof.mp h this
  exact Consistency this
\end{lstlisting}

\hypertarget{hybridrenamebound.lean}{%
\subsection{Hybrid/RenameBound.lean}\label{hybridrenamebound.lean}}

\emph{146 lines.}

\vspace{0.5\baselineskip}
\noindent\textcolor{green!40!black}{\textbf{Lean 4 source}}\par\vspace{0.25\baselineskip}
\lstinputlisting[style=leanbox]{lean-listings/Hybrid-RenameBound.lean}
\vspace{0.5\baselineskip}

\hypertarget{hybridformcountable.lean}{%
\subsection{Hybrid/FormCountable.lean}\label{hybridformcountable.lean}}

\emph{220 lines.}

\vspace{0.5\baselineskip}
\noindent\textcolor{green!40!black}{\textbf{Lean 4 source}}\par\vspace{0.25\baselineskip}
\lstinputlisting[style=leanbox]{lean-listings/Hybrid-FormCountable.lean}
\vspace{0.5\baselineskip}

\hypertarget{hybridlindenbaum.lean}{%
\subsection{Hybrid/Lindenbaum.lean}\label{hybridlindenbaum.lean}}

\emph{551 lines.}

\noindent\textcolor{green!40!black}{\textbf{Lean 4 source (lines 1--400)}}\par\vspace{0.25\baselineskip}
\lstinputlisting[style=leanbox,firstline=1,lastline=400]{lean-listings/Hybrid-Lindenbaum.lean}

\noindent\textcolor{green!40!black}{\textbf{Lean 4 source (lines 401--551)}}\par\vspace{0.25\baselineskip}
\lstinputlisting[style=leanbox,firstline=401,lastline=551]{lean-listings/Hybrid-Lindenbaum.lean}

\hypertarget{hybridlanguageextension.lean}{%
\subsection{Hybrid/LanguageExtension.lean}\label{hybridlanguageextension.lean}}

\emph{975 lines.}

\noindent\textcolor{green!40!black}{\textbf{Lean 4 source (lines 1--400)}}\par\vspace{0.25\baselineskip}
\lstinputlisting[style=leanbox,firstline=1,lastline=400]{lean-listings/Hybrid-LanguageExtension.lean}

\noindent\textcolor{green!40!black}{\textbf{Lean 4 source (lines 401--800)}}\par\vspace{0.25\baselineskip}
\lstinputlisting[style=leanbox,firstline=401,lastline=800]{lean-listings/Hybrid-LanguageExtension.lean}

\noindent\textcolor{green!40!black}{\textbf{Lean 4 source (lines 801--975)}}\par\vspace{0.25\baselineskip}
\lstinputlisting[style=leanbox,firstline=801,lastline=975]{lean-listings/Hybrid-LanguageExtension.lean}

\hypertarget{hybridexistencelemma.lean}{%
\subsection{Hybrid/ExistenceLemma.lean}\label{hybridexistencelemma.lean}}

\emph{183 lines.}

\vspace{0.5\baselineskip}
\noindent\textcolor{green!40!black}{\textbf{Lean 4 source}}\par\vspace{0.25\baselineskip}
\lstinputlisting[style=leanbox]{lean-listings/Hybrid-ExistenceLemma.lean}
\vspace{0.5\baselineskip}

\hypertarget{hybridcompletedmodel.lean}{%
\subsection{Hybrid/CompletedModel.lean}\label{hybridcompletedmodel.lean}}

\emph{954 lines.}

\noindent\textcolor{green!40!black}{\textbf{Lean 4 source (lines 1--400)}}\par\vspace{0.25\baselineskip}
\lstinputlisting[style=leanbox,firstline=1,lastline=400]{lean-listings/Hybrid-CompletedModel.lean}

\noindent\textcolor{green!40!black}{\textbf{Lean 4 source (lines 401--800)}}\par\vspace{0.25\baselineskip}
\lstinputlisting[style=leanbox,firstline=401,lastline=800]{lean-listings/Hybrid-CompletedModel.lean}

\noindent\textcolor{green!40!black}{\textbf{Lean 4 source (lines 801--954)}}\par\vspace{0.25\baselineskip}
\lstinputlisting[style=leanbox,firstline=801,lastline=954]{lean-listings/Hybrid-CompletedModel.lean}

\hypertarget{hybridcompleteness.lean}{%
\subsection{Hybrid/Completeness.lean}\label{hybridcompleteness.lean}}

\emph{109 lines.}

\vspace{0.5\baselineskip}
\noindent\textcolor{green!40!black}{\textbf{Lean 4 source}}\par\vspace{0.25\baselineskip}
\lstinputlisting[style=leanbox]{lean-listings/Hybrid-Completeness.lean}
\vspace{0.5\baselineskip}

\end{document}